\documentclass[a4paper,11 pt]{article}

\renewcommand{\baselinestretch}{2}

\usepackage[T1]{fontenc}
\usepackage[latin1]{inputenc}
\usepackage[font=small]{caption}
\usepackage{latexsym} %Simbolos adicionales de LaTeX
\usepackage{amssymb}
\usepackage{hyperref}
\usepackage{rotating}
\usepackage{fancyvrb}
\usepackage{color}
\usepackage{xcolor}
\usepackage{listings}
\usepackage{colortbl}

\definecolor{GGG}{RGB}{213,230,255}
\definecolor{GG}{RGB}{220,220,220}

\usepackage[T1]{fontenc}
\usepackage[latin1]{inputenc}
\usepackage[english]{babel}
\usepackage{latexsym} %Simbolos adicionales de LaTeX
\usepackage{amssymb}
\usepackage{amsmath}
\usepackage{array,longtable}
\usepackage{bbding} % paquete para estrellas de simbolos
\usepackage{mathpazo} %Font type: Palatino

\usepackage{apacite}
\usepackage[longnamesfirst]{natbib} %For BibTeX

\usepackage{amsthm}
\usepackage{xcolor}
\usepackage{multirow}

\usepackage{slashbox,booktabs,amsmath}

\usepackage{tikz}
\usepackage{algorithmic}
\usepackage{longtable}
\usepackage{arydshln}

\newcolumntype{P}[1]{>{\centering\arraybackslash}p{#1}}

\usepackage{natbib} %For BibTeX
\usepackage{setspace}
\allowdisplaybreaks

\usepackage[top=2.5cm,bottom=2.5cm,right=2.5cm,left=2.5cm,headsep=0.4in,includehead]{geometry}

\usepackage{lscape}
\usepackage{graphicx}
\usepackage{tikz}
\usetikzlibrary{spy}
\newlength\imagewidth
\newlength\imagescale

\theoremstyle{plain}

\newtheorem{lemma}{Lemma}[section]
\newtheorem{algorithm}{Algorithm}[section]

\theoremstyle{definition}

\newtheorem*{theorem*}{Theorem}

\theoremstyle{remark}

\newcommand{\reviewtimetoday}[2]{
}
\reviewtimetoday{\today}{Draft Version v.0.1}

\begin{document}
\title{\textbf{On systems of quotas based on bankruptcy with a priori unions: estimating random arrival-style rules % estimating the $RA$-random arrival rule
}}

\author{A. Saavedra-Nieves$^{\ast 1}$, P. Saavedra-Nieves$^1$}

\date{\footnotesize \emph{
		$^{\ast}$ \underline{Corresponding author}.
		$^1$ Departamento de Estatística, Análise Matemático e Optimización, Universidade de Santiago de Compostela.\texttt{alejandro.saavedra.nieves@usc.es, paula.saavedra@usc.es}
}}

\maketitle
\onehalfspacing
%\doublespacing
%\thispagestyle{empty}
%	\date{\today}
\renewcommand{\baselinestretch}{1.4}
\abstract{
{This paper addresses a sampling procedure for estimating extensions of the random arrival rule to those bankruptcy situations where there exist a priori unions. It is based on simple random sampling with replacement and it adapts an estimation method of the Owen value for transferable utility games with a priori unions, especially useful when the set of involved agents is sufficiently large. We analyse the theoretical statistical properties of the resulting estimator as well as we provide some bounds for the incurred error. Its performance is evaluated on two well-studied examples in literature where this allocation rule can be exactly obtained. Finally, we apply this sampling method to provide a new quota system for the milk market in Galicia (Spain) to check the role of different territorial structures when they are taken as a priori unions. The resulting quotas estimator is also compared with two classical rules in bankruptcy literature.}
 }
	
	$ $

	\setcounter{page}{1}
	\vspace{0.5cm}
	
	%{\bf JEL classification:} C44, C71
\tableofcontents

\section{Introduction}

Galicia is a Spanish autonomous community located in the Northwest of the Iberian Peninsula and in the Southwest of Europe with a remarkable dairy sector. In fact, it is the eighth dairy region in Europe and the leading dairy power in Spain employing around $25000$ people. However, Spanish National Institute of Statistics (INE)\footnote{Spanish National Institute of Statistics website: \url{https://www.ine.es}} indicates that the gross added value of the dairy activity in Galicia is barely 10\% of that of the Spanish, when the community has 40\% of the national milk and 60\% of the producers. Nowadays, just over 7500 farms produce milk in Galicia, a tenth of those that were still active in the early 1990s, when milk quota regime in the European Union (EU) began to be rigorously implemented in order to overcome the surpluses when the production of milk far outstripped the demand.

	\begin{figure}[h!]
	$\hspace{1.1cm}$\includegraphics[scale=.05]{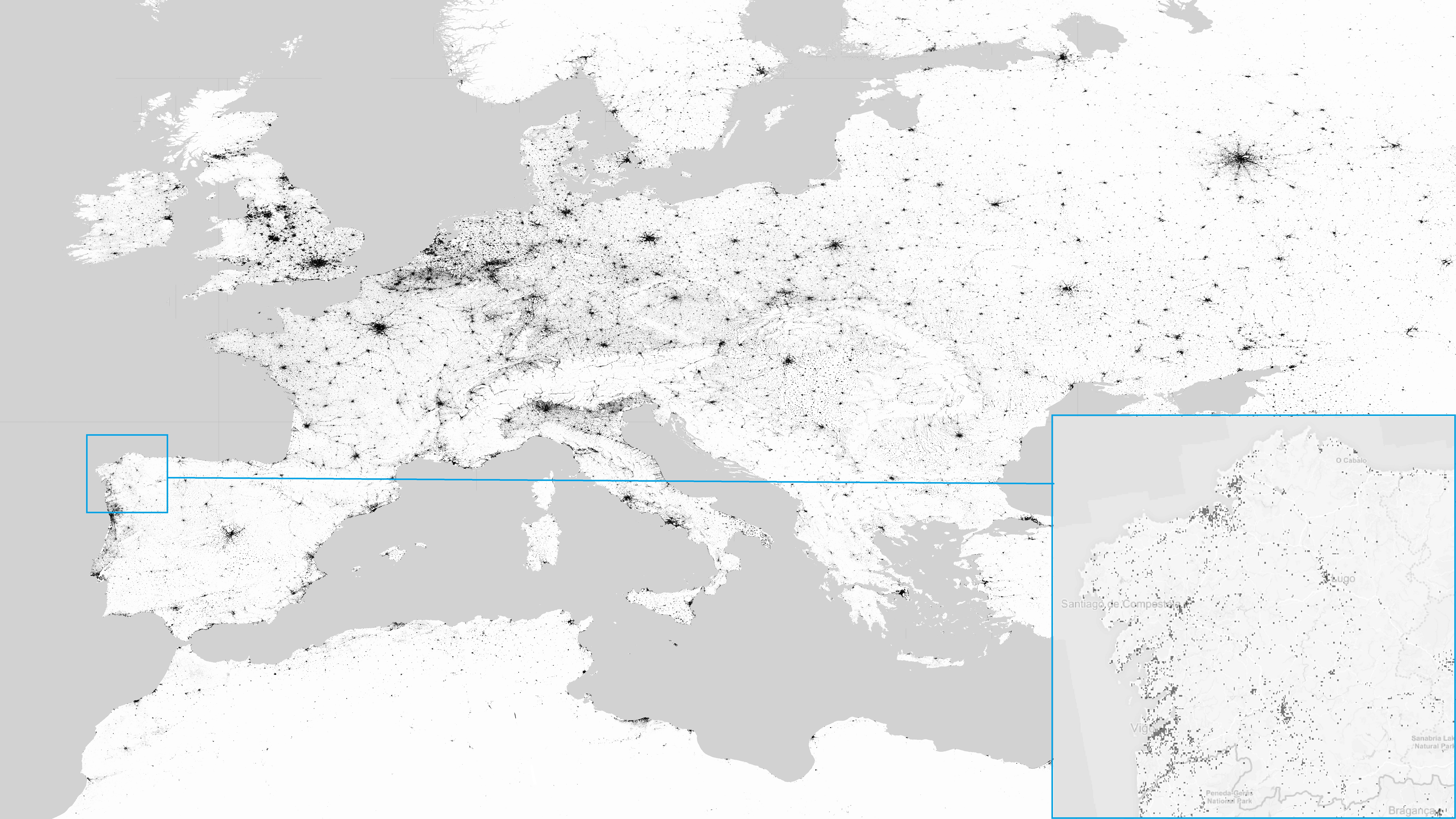}\\
		\caption{Geographical location of Galicia in Europe using a map that represents the population concentration.}\label{map33}
	\end{figure}

Those changes in the Common Agriculture Policy of the EU were devoted to help those dairy farmers from specially vulnerable areas providing EU producers more facilities in order to respond to an increasing demand of milk, by balancing price and production of milk. The end of the milk quotas regime in April 2015 caused an increase in production which derives into new problems in the dairy sector. The major consequence was that the milk price per liter of milk reduced. Mainly, because the EU was unable to predict the brutal production increase of countries such as Ireland and The Netherlands. Since several proposed initiatives were not able to solve the problem of milk prices, new measures after the abolition of these quotas such as a new Community policy or the establishment of a new upper-bound of the milk production could be addressed.

\cite{saavedrasaavedra} consider the problem of providing a new Galician system of milk quotas for the 190 councils that are milk producers. The main idea is that the the milk prices will increase if we assume that the aggregate of the Galician milk production has to decrease with respect to April 2015 (last official regulation of the EU milk quotas that is summarized in Appendix A). Although this solution can not be popular among milk producers in a first step, their profits will be perceived in the long-term. Mathematically, the previous situation is treated as a bankruptcy problem.

Bankruptcy problems have taken remarkable interest over the years due to their multiple applications. Their name refers to a usual problem in economics as the bankruptcy of a company. The existence of several creditors that claim a portion of a total estate is assumed. The main objective is to establish how we must divide the resource among all those agents who have claims on it. Of course, this is also the case for Galician milk problem. \cite{o1982problem} and \cite{aumann1985game} introduced first bankruptcy problems. A complete review on this topic can be seen in \cite{moulin2002axiomatic} or \cite{thomson2003axiomatic}. Furthermore, bankruptcy problems have been also considered from a game theoretic perspective in \cite{curiel1987bankruptcy}. Remark that any multi-agent allocation problem under cooperation can be studied as a bankruptcy case. It is only necessary the existence of an estate to be allocated where each agent claims a portion of the total. Therefore, the definition of rules for distributing the available resources is not a minor question in bankruptcy situations. Several alternatives of allocation procedures in bankruptcy were characterized in \cite{thomson2003axiomatic}. Three of the most common choices are the proportional rule, that assigns proportionally to the claims, the Talmud rule introduced by \cite{aumann1985game}, and the random arrival rule \citep{o1982problem} whose exact computation is a hard task. In fact, \cite{saavedrasaavedra} propose a method based on simple random sampling with replacement to approximate it.

Another interesting approach on bankruptcy problems was introduced in \cite{borm2005constrained}. The existence of a system of a priori unions on the set of agents involved incorporates additional information on the possibilities of their cooperation. Therefore, \cite{borm2005constrained} describe a two-stage procedure that extends bankruptcy rules when there exist a priori unions structures. Notice that \cite{borm2005constrained} prove that the extension of the random arrival rule coincides with the Owen value for the associated bankruptcy game with a priori unions \citep{Owen1975}. The main goal of this work is to propose a general procedure also based on simple random sampling with replacement for estimating the extension of the random arrival rule to this setting. Remark that its computation is again specially difficult when the number of involved agents increases. This estimation method will be analysed in terms of the incurred error by providing a theoretical bound and studying the statistical properties of the estimator. \vspace{-.9cm}

	\begin{figure}[h!]
\hspace{-1cm}\includegraphics[scale=.5]{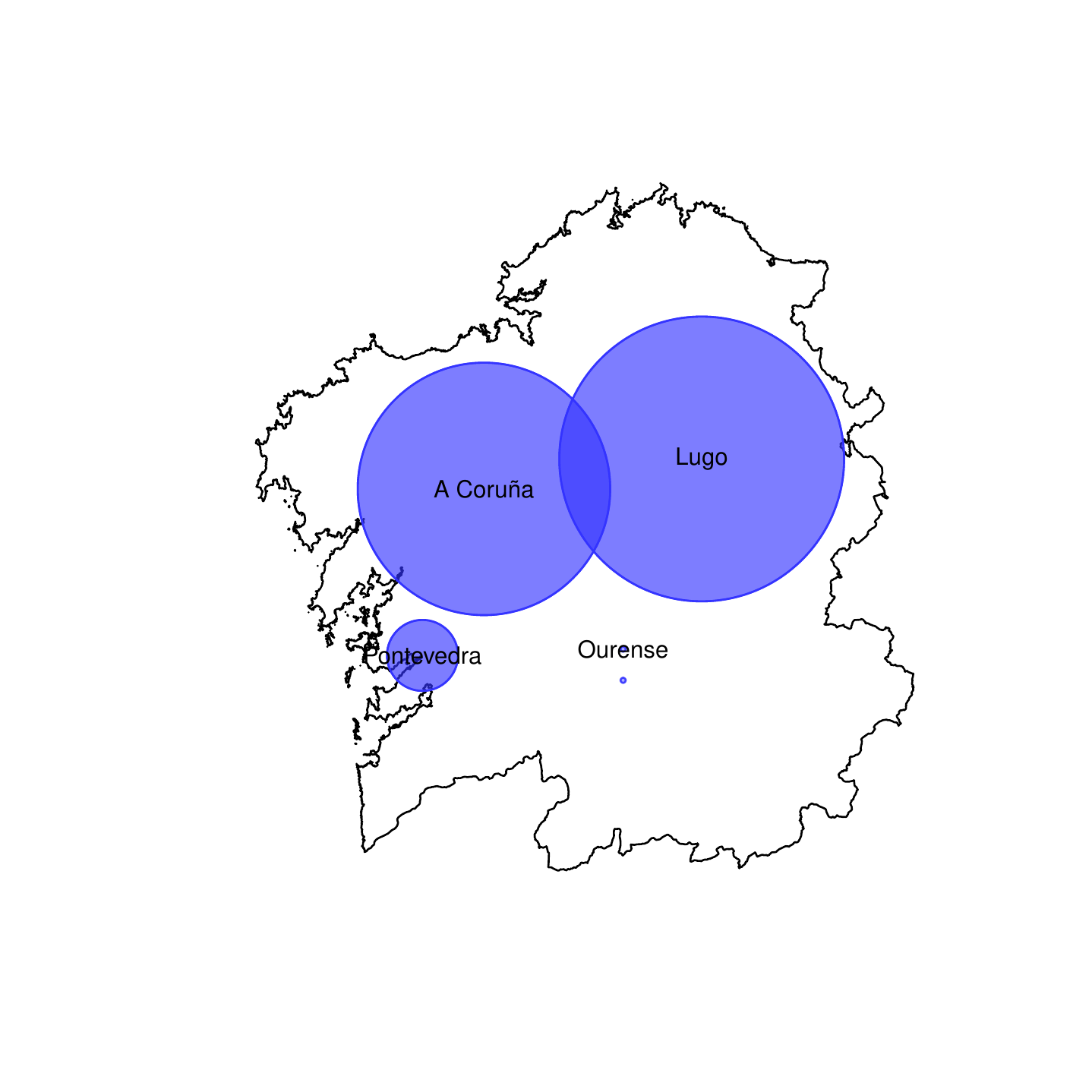}\hspace{-2.2cm}\includegraphics[scale=.5]{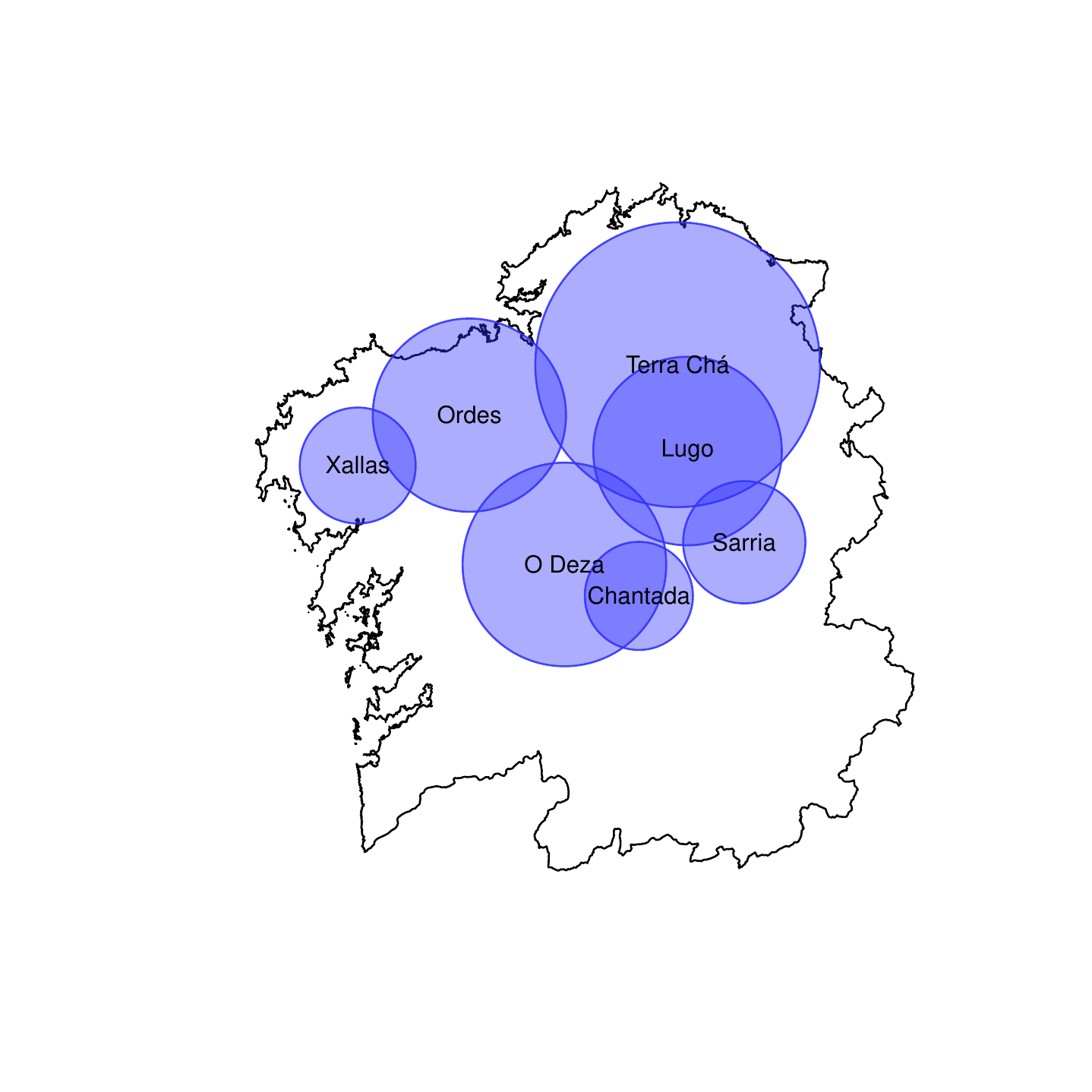}\hspace{-2.2cm}\includegraphics[scale=.5]{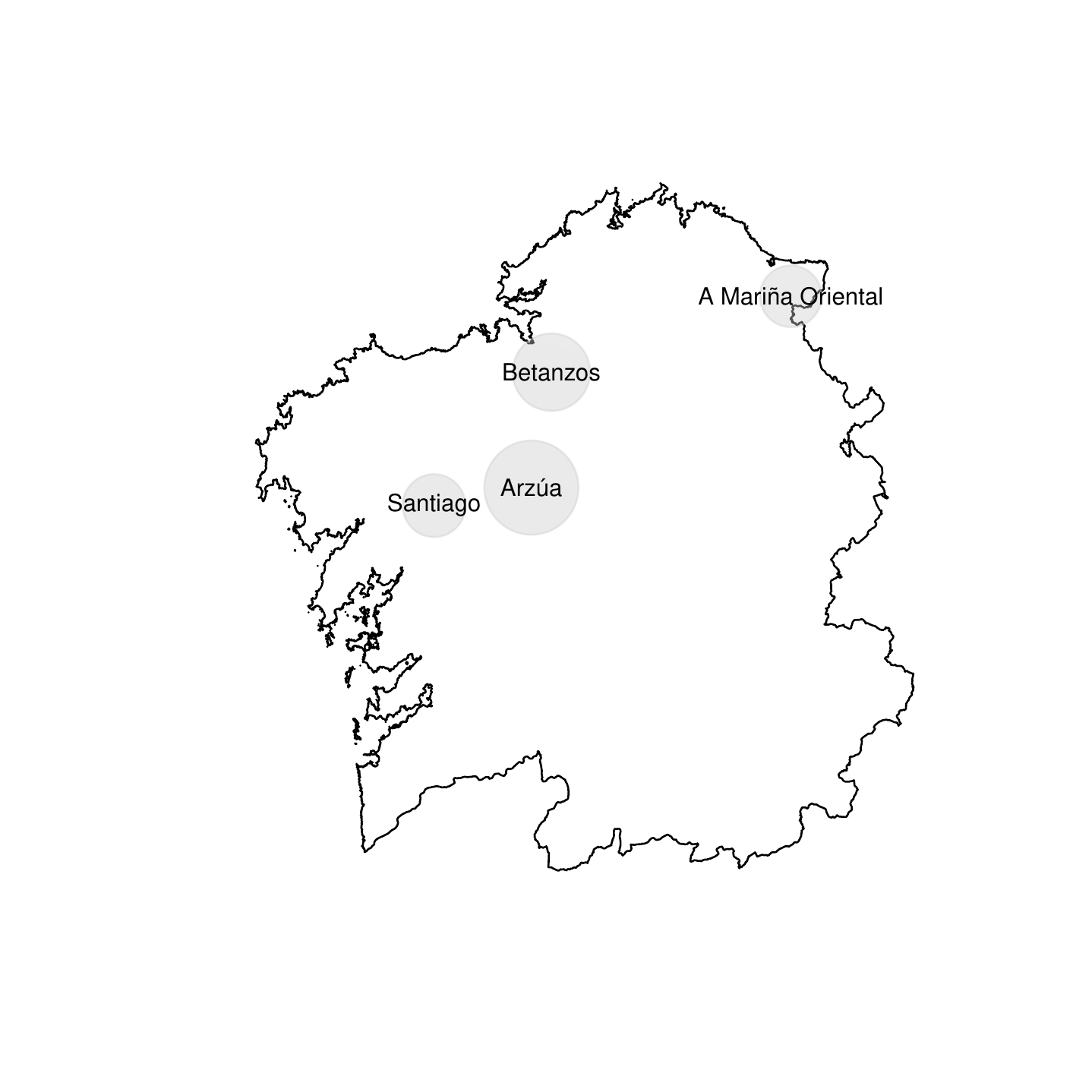}\vspace{-1.6cm}\\
	\caption{Bubble plots of the percentaged of aggregated milk quotas in the period 2014-2015 by provinces (left) and counties (center and right).
	}\label{map11}
\end{figure}

\begin{figure}[h!]
	 	$\hspace{.35cm}$\includegraphics[scale=.45]{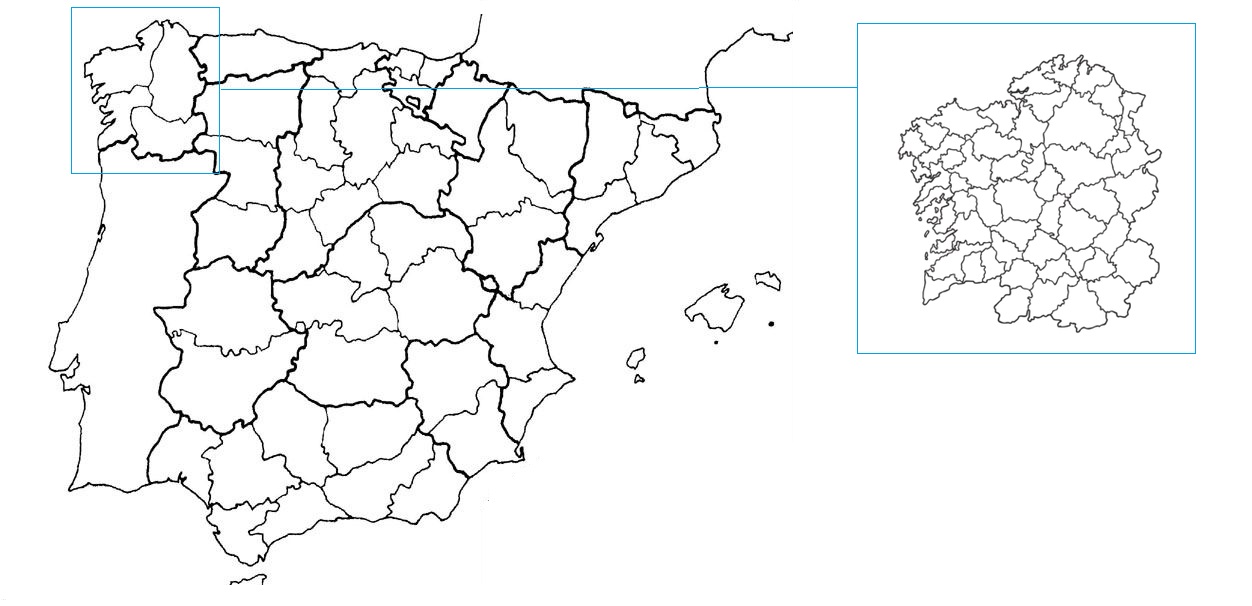}
	 	\caption{Geographical location of Galicia in the Northwest of the Iberian Peninsula. Administrative division in provinces of Spain (left) and in counties of Galicia (right).}\label{map1}
	 \end{figure}

As an autonomous community, Galicia is a territorial entity which, within the Spanish constitutional legal system, is endowed with autonomy, with its own institutions and representatives and certain legislative, executive and administrative powers. Moreover, it is also considered a historical nationality because of its collective, linguistic and cultural identity which is is different from the rest of Spain. Its territory, characterized by a very remarkable rural dispersion, is organized in four provinces (A Coruña, Lugo, Ourense and Pontevedra) which are made up of a total of 315 councils (in 2015). Map of European Space Agency\footnote{European Space Agency website: \url{https://www.esa.int/}} in Figure \ref{map33} shows the big dispersion of Galician population representing with black points all nonempty areas. The Decree 65/1997 of the Galician Government\footnote{Galician Government website: \url{https://www.xunta.gal/portada}} grouped these 315 councils into 53 counties that are represented in Figure \ref{map1}. The main objective of counties creation was to encourage different kinds of plans to undertake a structuring and revitalization of its territory overcoming the localisms that have always been the main
obstacle to the articulation of Galicia and, therefore, to achieve a real economic development. In this way, the rural areas would be strengthened combating the existing big problem of depopulation in Galicia where most of population is elderly people. Although some economists raise the need to prioritize the use of land over property rights and to organize services using the existing division in counties as the main measures to revitalize the rural areas, most of legislative initiatives and the strategies of territorial management and cooperation implanted during the last twenty years had not the desired effect. As an application of the methodology proposed in this work, we will consider the territorial divisions of Galicia in provinces and also in counties as a priori unions to check if these administrative organizations play a significant role in the milk conflict in Galicia. Figure \ref{map11} shows the the bubble plots of the percentages for aggregated milk quotas in the period 2014-2015 by provinces (left) and by counties that overcome 5\% (center) and 3\% (left) of the milk production, respectively. Tables \ref{sumary1} and \ref{sumary2} in Appendix A contains the exact values of these percentages. Concretely, Table \ref{sumary1} contains the aggregated milk quotas in the period 2014-2015 by provinces as well as the corresponding percentages.  Therefore, from the three bankruptcy rules cited above, several systems of Galician milk quotas will be provided considering these a priori unions. Results will be compared to the obtained ones in \cite{saavedrasaavedra} where territorial divisions were not taken into account.

This paper is organized as follows. Section \ref{sec2} reviews the bankruptcy problem formally including the extension of classical bankruptcy rules when there exist a priori unions. In the specific case of the extension of the random arrival rule to this setting, the proposed sampling procedure for its estimation when the number of involved agents increases is described in Section \ref{sec3}. Additionally, its statistical properties are studied and several theoretical results on bounding the error are provided. Then, the performance of this proposal is also evaluated on two well-known real examples in literature where the allocation rule can be exactly determined. Section \ref{sec5} contains the allocations of the extensions of the classical bankruptcy rules for each council in Galicia when the milk quota conflict is modeled as a bankruptcy problem.  Section \ref{sec6} presents the conclusions of this work. Finally, two appendices are included. Appendix A shows the milk quotas for $190$ councils in Galicia in the period 2014-2015 and their distribution in provinces and counties. Finally, Appendix B contains the new milk quota systems given by the estimations of the extensions of the random arrival rule and the classical Talmud and proportional rules when provinces and counties respectively determine systems of a priori unions. %Finally, Appendix C depicts the R code required for approximating the extension of the random arrival rule.

\section{Some background in bankruptcy}\label{sec2}

A \emph{bankruptcy problem} is a multi-agent situation in which agents claim a portion of a good larger than the one available. Formally, these situations were initially analysed in \cite{o1982problem}. A \emph{bankruptcy problem} with a set of claimants $N=\{1,\dots,n\}$ is given by
	$(N,c,E)$, where:
\begin{itemize}
\item 	 $E\in \mathbb{R}_{+}$ denotes the total amount to be divided, that is said to be the \emph{estate}; and 
\item $c\in \mathbb{R}^n$  is the vector of claims, where $c_i$ is the $i$'s \emph{claim}, with $i\in N$, and such that $0\leq E\leq \underset{j\in N}{\sum}c_j$.
\end{itemize}	
The set of bankruptcy problems with set of claimants $N$ is denoted by $B^N$.  \cite{o1982problem} also analyse these problems from a game-theoretical perspective.  A \emph{TU-game}  \citep{GonzalezDiaz2010} is formally given by a pair $(N,v)$, where  $N$ denotes the finite set of agents and $v : 2^N \longrightarrow \mathbb{R}$ is a map satisfying $v(\emptyset) = 0$. Thus, a \emph{bankruptcy game}  is a TU-game $(N,v)$ associated to each $(N,c,E)\in B^N$ and given, for each $S\subseteq N$, by  \begin{equation}\label{bank_game}v(S)=\max\bigg\{0,E-\sum_{i\notin S}c_i\bigg\}.\end{equation}%\end{definition}
For each $S\subseteq N$, $v(S)$ is the portion of $E$ that remains when agents in $N\setminus S$ received their claims. 

The definition of procedures for dividing the estate among the claimants is fundamental. A \emph{division rule} $f$ for a bankruptcy problem $(N,c,E)\in B^N$ is a function that assigns a vector $f (N,c,E)\in \mathbb{R}^n$ such that $\sum_{i\in N}f_i(N,c,E)=E$, and $0\leq f_i(N,c,E)\leq c_i$, for each $i\in N$. In this sense, several division rules have been introduced in  bankruptcy literature. 

Take $(N,c,E)\in B^N$ a bankruptcy problem. Among others, the \textit{Talmud rule} $T=(T_i)_{i\in N}$ assigns to each claimant $i\in N$, \[T_i(N,c,E)= \left\{ \begin{array}{lcc}
\min\left\{\frac{c_i}{2},\lambda\right\}, &   if  & E\leq \underset{l\in N}{\sum}\frac{c_l}{2},  \\
c_i-\min\left\{\frac{c_i}{2},\lambda\right\}, &  if & E\geq \underset{l\in N}{\sum}\frac{c_l}{2},
\end{array}
\right.\]
being $\lambda$ such that $\underset{l\in N}{\sum}T_l(N,c,E)=E$. \cite{aumann1985game} prove that the Talmud rule coincides with the nucleolus  \citep{schmeidler1969nucleolus} of the corresponding bankruptcy game. %We can say that the nucleolus is that allocation that minimizes the dissatisfaction of the players belonging to the less favoured coalitions.

The \emph{proportional rule} assigns to each $i\in N$, the part of $E$ corresponding to the weight given by its claim. Formally, it is given by $P=(P_i)_{i\in N}$, where
\[P_i(N,c,E)=\frac{c_i}{\underset{l\in N}{\sum}c_l}E, \mbox{ for each claimant }i\in N.\]

%
%Hereafter, we will focus on the random arrival rule and the computational problems that arise for its calculation. 
Let $\Pi(N)$ be the set of all permutations of $N$. For each $\pi\in \Pi(N)$, $P^\pi_i$ is the set of predecessors of $i$ under $\pi$. The \emph{random arrival rule} \citep{o1982problem}  assigns to each agent $i$ in $N$,\vspace*{-0.1 cm}
\begin{equation}\label{def_ra_rule}RA_i(N,c,E)=\frac{1}{n!}\underset{\sigma\in \Pi(N)}{\sum} \min\bigg\{c_i,\max\big\{0,E-\underset{j\in P_i^\sigma}{\sum}c_j\big\}\bigg\}.\vspace*{-0.1 cm}\end{equation}

\noindent This rule is obtained as the expected value of the marginal contributions for each $i\in N$. Among others, \cite{saavedrasaavedra} focus on the computational problem in determining the random arrival rule for those bankruptcy situations with large set of players.  To this aim, the sampling methodologies for estimating  the Shapley value for a general TU-game \citep{Castroetal2009,puerto2006teoria} since that the allocation given by the random arrival rule coincides with the Shapley value of the associated bankruptcy game.\footnote{{The Shapley value \citep{Shapley1953} of a TU-game $(N,v)$ is given by
		\[\Phi_i(N,v)=\sum_{S\subseteq N\setminus \{i\}}{\frac {|S|!\;(|N|-|S|-1)!}{|N|!}}(v(S\cup \{i\})-v(S)),\mbox{ for all }i\in N.\vspace*{-0.1 cm}\]
}} 

The existence of a system of a priori unions on the set of agents involved incorporate additional information about the possibilities of cooperation. Thus, bankruptcy problems can be also extended in this direction. A \emph{bankruptcy problem with a priori unions} will be denoted by $(N, E, c,P)$, being
$(N, c,E)$ a bankruptcy problem and $P = \{P_1,\dots,P_m\}$ is a partition of the set of players, being $M=\{1,\dots,m\}$. The class of all bankruptcy problems with a priori unions  will be denoted by $BU^N$. Thus, if $(N, c,E,P)\in BU^N$ is a bankruptcy problem with unions, we can define the corresponding bankruptcy problem among the unions $(M, c^P,E)$, the so-called quotient problem, where $c^P \in \mathbb{R}^m$ denotes the vector of total claims of the unions, so $c^P_k=\sum_{i\in P_k}c_i$, for each $k\in \{1,\dots,m\}$.

These problems can be also described by using cooperative game theory or, more precisely, TU-games with a priori unions. Recall that they are given by a triple $(N, v,P)$, where $(N, v)$ denotes TU-game and $P$ a partition of the set of players. For $(N, v,P)$, we define the associated TU-game for the unions, the \emph{quotient game}, whose characteristic function $v^U$ is given by $v^U(L) = v(\cup_{k\in L}P_k)$, for all $L\subseteq M$.

Analogously, the definition of bankruptcy rules under the existence of a structure of a priori unions has received attention in literature (see, for instance, \citealp{borm2005constrained}).  Formally, a \emph{division rule} $f^U$ for a bankruptcy problem with a priori unions $(N,c,E,P)\in BU^N$ denotes a function that assigns the vector $f^U(N,c,E,P)\in \mathbb{R}^n$ such that $\sum_{i\in N}f^U_i(N,c,E,P)=E$, and $0\leq f^U_i(N,c,E,P)\leq c_i$, for each $i\in N$. In this sense, several division rules have been introduced in  bankruptcy literature. 

For dividing the estate, \cite{borm2005constrained} initially describe a two-stage procedure that extends bankruptcy rules to  rules for bankruptcy situations with a priori unions. The idea that bases that procedure refers to firstly divide the estate among the unions and secondly, the allocation of each union is divided among the claimants belonging to the union. Formally, take $f:B^N\longrightarrow\mathbb{R}^N$ a bankruptcy rule and let $(N,c,E,P)\in BU^N$ be a bankruptcy rule with a priori unions. The steps of the Borm et al.' procedure to obtain $\overline{f}:BU^N\longrightarrow\mathbb{R}^N$ as an extension of $f$ to those situations with a priori unions are the following ones:
\begin{itemize}
\item First, define $E_k^f=f_k(M,c^P,E)$, for all $k\in M$ and, secondly, 
\item Do $\overline{f}_i(N,c,E,P)=f_i(P_k,(c_j)_{j\in P_k},E_k^f)$ for each $i\in P_k$ and for all $k\in M$.
\end{itemize}

Secondly, \cite{borm2005constrained} also use another two-step procedure for extending bankruptcy rules to those bankruptcy situations with a priori unions, based on the random arrival rule. Take $f$ a bankruptcy rule and  $(N,c,E,P)\in BU^N$. Then, the $f$-random arrival rule for $(N,c,E,P)$ is given by
\[RA_i^f(N,c,E,P)=\frac{1}{m!}\sum_{\sigma\in \Pi(M)}f_i(P_k,(c_j)_{j\in P_k},E_{\sigma}),\]
for all $i\in P_k$, where $E_{\sigma} = \max\{0,E-\sum_{l\in M,\sigma^{-1}(l)<\sigma^{-1}(k)}c_l^P\}$. Even more, \cite{borm2005constrained} prove that $RA^{RA}(N,c,E,P)=O(N,v,P)$, that is, it is the Owen value of the bankruptcy game. Notice that the \emph{Owen value} \citep{Owen1977}, that extends the Shapley value for TU-games with a priori unions, is defined by\begin{equation}\label{theOV}
O_i(N,v,P)=\sum_ {R\subseteq P\setminus P_{(i)}}\sum_{S\subseteq P_{(i)}\setminus \{i\}}\frac{s!(p_i-s-1)!r!(m-r-1)!}{p_i!m!}\big(v(\underset{P_l \in R}{\cup}P_l\cup S\cup\{i\})-v(\underset{P_l \in R}{\cup}P_l\cup S)\big),\end{equation}for every $i\in N$ and every $(N,v,P)\in U^N$.

The Owen value can  be alternatively formulated in
terms of permutations. A permutation ${\sigma}\in \Pi(N)$ is said to be \emph{compatible with a coalition structure} $P$ if %the elements of each class of $P$ are not separated by ${\sigma}$. Formally, if $\Pi_P(N)$ denotes the set of all permutations of $N$ which are compatible with $P$, then $\sigma\in\Pi_P(N)$ if and only if for all $i,j,k\in N$ 
it holds that $P_{(i)}=P_{(j)}\mbox{ and } {\sigma}(i) < {\sigma}(k) < {\sigma}(j) \mbox{ implies that } P_{(k)}=P_{(i)}=P_{(j)}$. Therefore, the Owen value of a bankruptcy game with a priori unions $(N,v,P)$ can be rewritten,  for every $i\in N$, as
\begin{equation}
\label{the_owen_value}
O_{i}(N,v,P)=\frac{1}{|\Pi_P(N)|}\sum _{\sigma \in \Pi_P(N)}\min\bigg\{c_i,\max\big\{0,E-\underset{j\in P_i^\sigma}{\sum}c_j\big\}\bigg\},
\end{equation}
where $\Pi_P(N)$ denotes the set of  permutations of $N$ compatible with $P$.

\section{Estimating the $RA$-random arrival rule}\label{sec3}

The computation of the $RA$-random arrival rule for bankruptcy situations with a priori unions becomes a difficult task when the number of involved agents increases since the number of compatible permutations to be evaluated exponentially increases. In this section, we propose a general procedure for estimating this rule based on simple random sampling with replacement. It is an application of the  proposal in \cite{Saavedra2018a} for approximating the Owen value for general TU-games to bankruptcy problems with a priori unions. This methodology avoids the usage of the corresponding bankruptcy game.

Although the aim of this section is focused on the estimation of the $RA-$random arrival rule, it is important to mention that problems of its exact computation can arise in general for the $f-$random arrival rule family, being $f$ any of the classic bankruptcy rules. In the case of applying the Talmud rule or the proportional rule in one of the steps, difficulties would arise only in the step of using the random arrival rule. In order to address these difficulties, the sampling proposal for estimating the random arrival rule in \cite{saavedrasaavedra} can be used.

\subsection{The sampling algorithm}\label{samplingalgorithm}

Below, we formally describe the procedure for approximating the $RA$-random arrival rule for bankruptcy situations with a priori unions based on sampling techniques.

Let $(N,c,E,P)\in BU^N$ be a bankruptcy problem with a priori unions. The steps of the sampling proposal are described below.
\begin{enumerate}
	\item The population of the sampling procedure is the set of those permutations of $N$ compatible with $P$, i.e. $\Pi_P(N)$.
	\item The vector of parameters to be estimated is $RA^P = (RA_i^P)_{i\in N}$, where $RA^{P}_i$ denotes $RA^{RA}_i(N,c,E,P)$ for all $i\in N$.
	\item The characteristic to be studied in each sampling unit $\sigma \in \Pi_P(N)$ is the vector $(x(\sigma)_i)_{i\in N}$,
	where \vspace*{-0.12 cm}\begin{equation}\label{mc:bankruptcy}x(\sigma)_i= \min\bigg\{c_i,\max\big\{0,E-\underset{j\in P_i^\sigma}{\sum}c_j\big\}\bigg\},\mbox{ for all }i\in N.\vspace*{-0.12 cm}\end{equation}
	\item
	The sampling procedure takes each permutation $\sigma \in \Pi(N)$ with the same probability. By construction, we obtain a sample  with replacement  $\mathcal{S}=\{\sigma_1,\dots,\sigma_{\ell}\}$ of $\ell$  orders of $N$.
	
	\item The estimation of the $i^{th}$-component of $RA^{P}$, $RA^{P}_i$, is the mean of the
	marginal contributions vectors over the sample of permutations $\mathcal{S}$. Formally, we obtain a vector
	$\overline{RA}^P=(\overline{RA}^P_i)_{i\in N}$
	where, for each $i\in N$,  $\overline{RA}^P_i=\frac{1}{\ell}\sum_{\sigma\in \mathcal{S}}x(\sigma)_i$   approximates $RA_i$ being $\ell$ the sampling size.
\end{enumerate}

Algorithm \ref{pseudocode_RA} presents the pseudo code of this sampling procedure.
\begin{algorithm}\label{pseudocode_RA}
	Let $(N,c,E,P)\in BU^N$ be a bankruptcy problem with a priori unions.
	\begin{algorithmic}
		\STATE Set $\ell$. Do $cont=0$ and $\overline{RA}^P_i=0$, for all $i\in N$.
		\WHILE { $cont<\ell$}
		\STATE Take $\sigma \in \Pi_P(N)$ with probability $\frac{1}{n!}$.
		\FOR{$i\in N$}
		\STATE Calculate $P_i^{\sigma}$ and $x(\sigma)_i
		= \min\big\{c_i,\max\big\{0,E-\underset{j\in P_i^\sigma}{\sum}c_j\big\}\big\}.$
		\STATE Do $\overline{RA}^P_i=\overline{RA}^P_i+x(\sigma)_i$.
		\ENDFOR
		\STATE $cont=cont+1$.
		\ENDWHILE 	
		\STATE Finally, $\overline{RA}^P_i=\frac{\overline{RA}^P_i}{\ell}$ for each $i\in N$.
	\end{algorithmic}
\end{algorithm}

By construction, the problem of estimating the $RA$-random arrival rule for $i\in N$ corresponds to the approximation of the expected value of the marginal contributions $x(\sigma)_i$. Thus, we are dealing with a very common task in Statistics for which sampling, and more specifically, simple random sampling  becomes a useful. Below, we do a statistical analysis of the properties that our sampling procedure satisfies. Given an  agent $i\in N$, the estimator $\overline{RA}^P_i$ is unbiased:\vspace*{-0.075 cm}
$$\mathbb{E}(\overline{RA}^P_i) =\mathbb{E}\bigg(\frac{1}{\ell}\sum_{\sigma \in \mathcal{S}}x(\sigma)_i\bigg)=\mathbb{E}(x(\sigma)_i)=RA^{P}_i.\vspace*{-0.075 cm}$$
Besides,\vspace*{-0.075 cm}
$$\mbox{Var}(\overline{RA}^P_i) = \mbox{Var}\bigg(\frac{1}{\ell}\sum_{\sigma \in \mathcal{S}} x(\sigma)_i\bigg)=\frac{\theta^2}{\ell}\vspace*{-0.075 cm}$$
where $\theta^2$ denotes the variance of $x(\sigma)_i$ with respect to $RA^{P}_i$. Hence, taking into account that\vspace*{-0.075 cm}
\begin{equation*}
	MSE(\overline{RA}^P_i)=\mathbb{E}(\overline{RA}^P_i-RA^P_i)^2=(\mathbb{E}(\overline{RA}^P_i)-RA^P_i)^2 +\mbox{Var}(\overline{RA}^P_i),\vspace*{-0.09 cm}
\end{equation*}
and the unbiased character of $\overline{RA}^P_i$,  $MSE(\overline{RA}^P_i)=\mbox{Var}(\overline{RA}^P_i)$, that goes to zero when $\ell$ increases.

\subsection{Analysis of the error}

A fundamental issue in a problem as the one dealt with focuses on bounding the absolute  error in the estimation. Since this error is often not possible to be measured, a probabilistic bound on its value is theoretically provided instead. This means that the error is guaranteed to be not within a bound $\varepsilon$ with a certain probability $\alpha$ as maximum. Formally, this is equivalent to\vspace*{-0.2 cm}
\begin{equation*}
\mathbb{P}(|\overline{RA}^P_i-{RA}^{P}_i|\geq \varepsilon)\leq\alpha.\vspace*{-0.2 cm}
\end{equation*}
It is easy to check that the estimated value usually is a good approximation of the random arrival rule for bankruptcy for sufficiently large sampling sizes. In what follows, we state a collection of statistical results that can be useful for determining the required sample size.

As in \cite{saavedrasaavedra}, Tchebyshev's inequality helps with bounding the incurred estimation error in the approximation of the $RA$-random arrival rule, by using the variance of the marginal contributions. In most cases, this amount is not  explicitly determined and, for this reason, \textit{Popoviciu's inequality on variances} \citep{Popoviciu1935} solves this drawback. In this case, the range of the marginal contributions in bankruptcy is mandatory for analysing the error. Its value for a fixed $i\in N$, denoted by $w_i$, is given by \vspace*{-0.2 cm}
	$$w_i=\underset{\sigma,\sigma'\in \Pi(N)}{\max}\big\{x(\sigma)_i-x(\sigma')_i\big\}.\vspace*{-0.15 cm}$$
Notice that $0\leq x(\sigma)_i\leq c_i$ for a fixed agent $i$ since that, by Expression (\ref{mc:bankruptcy},\vspace*{-0.08 cm}
\[x(\sigma)_i= \left\{ \begin{array}{lcl}
c_i &   if  & E \geq \sum_{j\in P_i^{\sigma}}c_j+c_i; \\
 E - \sum_{j\in P_i^{\sigma}}c_j &  if & 0\leq E - \sum_{j\in P_i^{\sigma}}c_j\leq c_i; \\
 0 &  if  & E \leq \sum_{j\in P_i^{\sigma}}c_j.
\end{array}
\right.\vspace*{-0.07 cm}\]
Thus, it is bearable that $w_i=c_i$. Consequently, we write two usual results to bound the incurred error in the estimation of the random arrival rule directly in terms of the value of $c_i$. They can be considered as specific theoretical results for bankruptcy. By combining this inequality with the bound given by  \textit{Hoeffding's inequality} \citep{Hoeffding} the following lemma can be extracted. For more statistical details, see \cite{Saavedra2018a}.  It establishes a very general bound of the absolute error in terms of the claims of agents for bankruptcy problems with a priori unions.

\begin{lemma}
	\label{saavedra_cor2}
		\citep{saavedrasaavedra} Let $(N,c,E,P)\in BU^N$ be a bankruptcy problem with a priori unions and take $\varepsilon>0$ and $\alpha\in (0,1)$. Then,
		\[\ell\geq \min\bigg\{\frac{1}{4\alpha\varepsilon^2},\frac{\ln(2/\alpha)}{2\varepsilon^2}\bigg\}c_i^2\mbox{ implies that } \mathbb{P}(|\overline{RA}^P_i-RA^P_i|\geq \varepsilon)\leq \alpha.\]
		
\end{lemma}

Below, we remind some helpful ideas in the estimation of the $RA$-random arrival rule for bankruptcy problems with a priori unions. For the usual values of $\alpha$ ($\alpha=0.1$, $\alpha=0.05$ or $\alpha=0.01$), \cite{Maleki2015} proves that Hoeffding's inequality requires a smaller sampling size than Chebyshev's inequality. In our setting,
\[\min\bigg\{\frac{1}{4\alpha\varepsilon^2},\frac{\ln(2/\alpha)}{2\varepsilon^2}\bigg\}c_i^2=\frac{\ln(2/\alpha)}{2\varepsilon^2}c_i^2.\]
%On the other hand,  the number of samples required to estimate the $RA$-random arrival value is
%\[O(p(n)ln(1/\alpha))\]
%for a given value $\varepsilon = \frac{1}{p(n)}$ with $p(n)$ a polynomial, and with a probability of $1-\alpha$. When $\alpha$ is exponentially small, the required amount of samples is polynomial.

\subsection{Algorithm performance on two numerical examples}\label{sec4}

In this section, we analyse how our sampling proposal performs in two examples extracted from the literature for which the exact $RA$-random arrival rule is obtained in a reasonable time. Following \cite{borm2005constrained}, it coincides with the Owen value of the corresponding bankruptcy game in (\ref{theOV}). We consider the examples used in \cite{saavedrasaavedra}, again because of the absence of other examples in the literature in which this rule is calculated for large sets of players. The main purpose of our methodology is to address these types of seemingly difficult problems from a computational point of view.

The results included in next sections have been performed computing our sampling proposal into the statistical software R \citep{Rsoftware} on a personal computer with Intel(R) Core(TM) i5-7400 and 8 GB of memory and with a single 3.00GHz CPU processor.

\subsubsection{The Pacific Gas and Electric Company}\label{sec41}

This example of bankruptcy is extracted from \cite{borm2005constrained}. It describes the situation that arises when the American company \emph{Pacific Gas and Electric Company} declares bankruptcy. The creditors can be organized according to the nature of their claims: bank bonds, power purchases and gas purchases. So, the system of a priori unions is given by $P=\{P_1,P_2,P_3\}$, where 
\begin{itemize}
	\item $P_1 = \{1, 3, 5, 6\}$, 
	\item $P_2 =\{2, 4, 7, 8, 9, 10, 11, 12, 13, 18, 19, 20\}$, and \item $P_3 = \{14, 15, 16, 17\}$.
\end{itemize}

First, Table \ref{claims_ra} depicts the nature of claims, the claims and the $RA$-random arrival rule as the Owen value of the bankruptcy game in (\ref{bank_game}).

\vspace*{0.4 cm}\begin{table}[ht]
	\begin{center}
		\resizebox{0.9\textwidth}{!}{	
			\begin{tabular}{|p{0.02\textwidth}||p{0.2\textwidth}|p{0.1\textwidth}|p{0.08\textwidth}|||p{0.03\textwidth}||p{0.2\textwidth}|p{0.1\textwidth}|p{0.08\textwidth}|}\hline
				$i$  & Nature of claims & $c_i$ & $RA^{RA}_i$ & $i$  &Nature of claims & $c_i$ & $RA^{RA}_i$  \tabularnewline\hline
				1&Bank bonds &2207.2500&161.4321&11&Power purchases&40.1472&6.6912\tabularnewline
				2&Power purchases &1966.0000&224.5992&12&Power purchases&40.1221&6.6870\tabularnewline
				3& Bank bonds&1302.1000&161.4321&13&Power purchases&32.8679&5.4780\tabularnewline
				4&Power purchases &1228.8000&224.5992&14&Gas purchases&29.5235&9.8412\tabularnewline
				5&Bank bonds&938.4610&147.0812&15&Gas purchases&28.2106&9.4035\tabularnewline
				6&Bank bonds&310.0000&42.3377&16&Gas purchases&24.7183&8.2395\tabularnewline
				7&Power purchases&57.9284&9.6547&17&Gas purchases&23.8495&7.9498\tabularnewline
				8&Power purchases&49.4526&8.2421&18&Power purchases&22.5765&3.7628\tabularnewline
				9&Power purchases&48.4006&8.0668&19&Power purchases&21.5061&3.5844\tabularnewline
				10&Power purchases&45.7064&7.6177&20&Power purchases&19.8002&3.3000\tabularnewline\hline
		\end{tabular}}
	\caption{Nature of claims, claims and the $RA$-random arrival rule for the bankruptcy problem with a priori unions $(N,c,E,P)$.}\label{claims_ra}
	\end{center}
\end{table}

Table \ref{claims_ra} displays the claims (in millions of dollars) for the set  of the different agents $N=\{1,\dots,20\}$ that are involved. The estate to be allocated among the creditors is equal to $E=1060$ millions of dollars. Table \ref{claims_ra} also displays the vector $RA^{RA}=(RA^{RA}_1,\dots,RA^{RA}_{20})$  provided by the $RA$-random arrival rule for this bankruptcy situation with a priori unions (of dimension $20$).

\vspace*{0.4 cm}\begin{table}[ht]
	\begin{center}
			\resizebox{0.7\textwidth}{!}{	
		\begin{tabular}{|p{0.1\textwidth}|p{0.13\textwidth}|p{0.13\textwidth}|p{0.13\textwidth}|p{0.13\textwidth}|p{0.13\textwidth}|}\hline
			$i$ & 1 & 2 & 3 & 4 & 5\tabularnewline\hline
			$\alpha=0.1$ & 2919013106 &2315794515& 1015831491&  904680854& 527673705\tabularnewline
			$\alpha=0.05$& 3594409142 &2851618912& 1250872766& 1114004294& 649765903\tabularnewline
			$\alpha=0.01$& 5162630175 &4095764632& 1796621706& 1600038269& 933255210\tabularnewline\hline\hline
			$i$& 6 & 7 & 8 & 9 & 10\tabularnewline\hline
			$\alpha=0.1$ &  57577975 &2010555 &1465250 &1403570& 1251661 \tabularnewline
			$\alpha=0.05$&  70900264 &2475753 &1804276 &1728326 &1541268\tabularnewline
			$\alpha=0.01$& 101833660 &3555911 &2591472 &2482384 &2213715\tabularnewline\hline\hline
			$i$& 11 & 12 & 13 & 14 & 15\tabularnewline\hline
			$\alpha=0.1$ &  965706 & 964495 &647257 &522240& 476822 \tabularnewline
			$\alpha=0.05$&  1189149& 1187658& 797018& 643075& 587149\tabularnewline
			$\alpha=0.01$& 1707967 &1705826& 1144752& 923644& 843318\tabularnewline\hline\hline
			$i$& 16 & 17 & 18 & 19 & 20\tabularnewline\hline
			$\alpha=0.1$ & 366077 &340793& 305385& 277113 &234896 \tabularnewline
			$\alpha=0.05$&  450779 &419645 &376044 &341231 &289245\tabularnewline
			$\alpha=0.01$&  647451 &602733& 540110& 490107 &415441\tabularnewline\hline
		\end{tabular}}
		\caption{Sampling sizes for estimating the $RA$-random arrival rule with $w_i=c_i$ for all $i\in N$.}\label{sampling_sizes}
	\end{center}
\end{table}

\vspace*{-0.2 cm}Below, the $RA$-random arrival rule is estimated by using our sampling proposal. We take $\varepsilon=0.1,0.05\mbox{ or }0.01$, or equivalently, $1000$ thousands, $50$ thousands or a thousand of dollars. They are values that can be naturally assumed. In this sense, we show in Table \ref{sampling_sizes} the minimum sampling sizes required to ensure that absolute errors smaller than or equal to $\varepsilon=0.05$ with probability at least $1-\alpha$.

First, we approximate the $RA$-random arrival rule with several sampling sizes and then, we do a comparison with the exact allocation. This analysis is done in terms of the incurred absolute error and the estimated variance. Both measures are reduced when $\ell$ increases.

\vspace*{0.25 cm}\begin{table}[ht]
	\begin{center}
			\resizebox{0.9\textwidth}{!}{	
		 \begin{tabular}{|p{0.02\textwidth}||p{0.12\textwidth}|p{0.1\textwidth}|p{0.125\textwidth}||p{0.12\textwidth}|p{0.1\textwidth}|p{0.13\textwidth}||p{0.125\textwidth}|p{0.1\textwidth}|p{0.13\textwidth}|}\hline
			 \multirow{2}{*}{$i$}& \multicolumn{3}{c||}{$\ell=10^3$}&\multicolumn{3}{c||}{$\ell=10^5$}&\multicolumn{3}{c|}{$\ell=10^7$}\tabularnewline\cline{2-10}
			  &  Abs. error & Err. th., $\alpha=0.01$ & Estimated variance & Abs. error & Err. th., $\alpha=0.01$ & Estimated variance & Abs. error & Err. th., $\alpha=0.01$  & Estimated variance \tabularnewline\hline
 1 & 12.9658 & 113.6071 & 135.1849 &0.0208& 11.3607 &1.2743&  0.1322& 1.1361&  $ 1.2719\cdot 10^{-2}$\\
2 & 1.5985 & 101.1900 & 152.9544 & 2.5888 &10.1190 & 1.5315 & 0.0838 & 1.0119     &  $1.5457\cdot 10^{-2}$\\
3 & 7.2830 & 67.0191 & 132.3190 &0.0247 & 6.7019 &1.2724 & 0.1248 & 0.6702    & $1.2719\cdot 10^{-2}$\\
4 & 16.0863 & 63.2463 & 145.4868&  0.7074 & 6.3246 & 1.5493 & 0.2322 & 0.6325 & $1.5473\cdot 10^{-2}$\\
5 & 7.0849 & 48.3026 & 104.6027 &1.6193 & 4.8303 &1.1077 & 0.1043 & 0.4830    & $1.0990\cdot 10^{-2}$\\
6 & 3.5450 & 15.9557 & 11.3958 & 0.2678 & 1.5956 & 0.1068 & 0.0091 & 0.1596   & $1.0630\cdot 10^{-3}$\\
7 & 0.1931 &  2.9816 & 0.4735 & 0.0311 & 0.2982 & 0.0046 & 0.0057 & 0.0298 & $4.6629\cdot 10^{-5}$\\
8 & 0.5769 &  2.5453 & 0.3203 & 0.0226 & 0.2545 & 0.0034 & 0.0024 & 0.0255 & $3.3974\cdot 10^{-5}$\\
9 & 0.4517 &  2.4912 & 0.3397 & 0.0148 & 0.2491 & 0.0032 & 0.0074 & 0.0249 & $3.2560\cdot 10^{-5}$\\
10 & 0.1066 &  2.3525 & 0.2934 & 0.0812 & 0.2353 & 0.0029 & 0.0064 & 0.0235 & $2.8996\cdot 10^{-5}$\\
11 & 0.0268 &  2.0664 & 0.2231 & 0.0043 & 0.2066 & 0.0022 & 0.0067 & 0.0207 & $2.2404\cdot 10^{-5}$\\
12 & 0.1471 &  2.0651 & 0.2196 & 0.0007 & 0.2065 & 0.0022 & 0.0024 & 0.0207 & $2.2365\cdot 10^{-5}$\\
13 & 0.3177 &  1.6917 & 0.1430 & 0.0090 & 0.1692 & 0.0015 & 0.0036 & 0.0169 & $1.5012\cdot 10^{-5}$\\
14 & 0.4330 &  1.5196 & 0.1978 & 0.0253 & 0.1520 & 0.0019 & 0.0028 & 0.0152 & $1.9367\cdot 10^{-5}$\\
15 & 0.4138 &  1.4520 & 0.1806 & 0.0242 & 0.1452 & 0.0018 & 0.0027 & 0.0145 & $1.7683\cdot 10^{-5}$\\
16 & 0.3625 &  1.2723 & 0.1386 & 0.0212 & 0.1272 & 0.0014 & 0.0024 & 0.0127 & $1.3576\cdot 10^{-5}$\\
17 & 0.3498 &  1.2275 & 0.1291 & 0.0204 & 0.1228 & 0.0013 & 0.0023 & 0.0123 & $1.2638\cdot 10^{-5}$\\
18 & 0.1731 &  1.1620 & 0.0682 & 0.0468 & 0.1162 & 0.0007 & 0.0020 & 0.0116 & $7.0821\cdot 10^{-6}$\\
19 & 0.1577 &  1.1069 & 0.0665 & 0.0009 & 0.1107 & 0.0006 & 0.0002 & 0.0111 & $6.4241\cdot 10^{-6}$\\
20 & 0.2508 &  1.0191 & 0.0511 & 0.0050 & 0.1019 & 0.0005 & 0.0003 & 0.0102 & $5.4447\cdot 10^{-6}$\\\hline
		\end{tabular}}
		\caption{Estimation of the $RA$-random arrival rule for $(N,c,E,P)$ for several values of $\ell$.}\label{est_bankruptcy}
	\end{center}
\end{table}

\newpage Those columns of Table \ref{est_bankruptcy} containing the absolute errors show that our sampling proposal correctly estimates the $RA$-random arrival rule in this example. These amounts are always  smaller than the ones provided by Lemma \ref{saavedra_cor2} for this particular estimation. To check this conjecture in general, we do a small simulation study  by using $\ell=10^4$.  Table \ref{100_simus} reflects the minimum, maximum and mean observed absolute errors in $1000$ estimations as well as the theoretical errors for $\alpha=0.01$. We conclude that the absolute errors are smaller than the theoretical ones.

\begin{table}[h!]
	\begin{center}
			\resizebox{0.78\textwidth}{!}{	
		\begin{tabular}{|p{0.24\textwidth}|p{0.12\textwidth}|p{0.12\textwidth}|p{0.12\textwidth}|p{0.12\textwidth}|p{0.12\textwidth}|}\hline
			$i$ & 1 & 2 & 3 & 4 & 5\tabularnewline\hline
			Theoretical, $\alpha=0.01$ & 35.9257 &31.9991& 21.1933& 20.0002& 15.2746\tabularnewline\hline
			Maximum &  9.7670 &11.1829 &10.1735 &11.8164&  8.1748\tabularnewline
			Average & 2.7228  &2.9959  &2.8935  &3.1117 & 2.5920\tabularnewline
			Minimum & 0.0663  &0.0605  &0.0198  &0.0166 & 0.0413\tabularnewline\hline\hline
			$i$& 6 & 7 & 8 & 9 & 10\tabularnewline\hline
			Theoretical, $\alpha=0.01$ &  5.0456& 0.9429& 0.8049& 0.7878& 0.7439 \tabularnewline\hline
		Maximum & 2.8388& 0.5812& 0.4714& 0.6357& 0.4632\tabularnewline
		Average & 0.9316& 0.1734& 0.1568& 0.1567& 0.1447\tabularnewline
		Minimum & 0.0037& 0.0019& 0.0033& 0.0032& 0.0076\tabularnewline\hline\hline
			$i$& 11 & 12 & 13 & 14 & 15\tabularnewline\hline
			Theoretical, $\alpha=0.01$ &  0.6534 &0.6530& 0.5350& 0.4805& 0.4592 \tabularnewline\hline
			Maximum & 0.4644& 0.3477& 0.3440& 0.2933& 0.2802\tabularnewline
			Average & 0.1319& 0.1173& 0.1071& 0.1138& 0.1088\tabularnewline
			Minimum & 0.0054& 0.0027& 0.0011& 0.0010& 0.0009\tabularnewline\hline\hline
			$i$& 16 & 17 & 18 & 19 & 20\tabularnewline\hline
			Theoretical, $\alpha=0.01$ & 0.4023& 0.3882& 0.3675& 0.3500& 0.3223\tabularnewline\hline
			Maximum &  0.2455 &0.2369& 0.2092& 0.2437& 0.2244\tabularnewline
			Average & 0.0953 &0.0919& 0.0638 &0.0691 &0.0645\tabularnewline
			Minimum & 0.0008& 0.0008 &0.0008& 0.0007& 0.0013\tabularnewline\hline
		\end{tabular}}
		\caption{Summary of the absolute errors in the $1000$ simulations.}\label{100_simus}
	\end{center}
\end{table}

\newpage

\subsubsection{A conflictive situation in university management}\label{sec42}

Along this section, the bankruptcy example in \cite{pulido2002game} is used to illustrate the usage of the sampling proposal. This situation, from a bankruptcy approach, is proposed for allocating  resources in a Spanish university through the distribution of the available money among the entities to finance the purchase of equipment.  Table \ref{claims_ra_ex2} describes the elements that characterize the problem: the system of unions $P$ to which each player belongs, the claims  for the set  of the agents $N=\{1,\dots,27\}$ and the corresponding allocation given by the $RA$-random arrival rule. Notice that although this example is considered without assuming the existence of a system of a priori unions, we take $P=\{P_1,P_2,P_3,P_4,P_5,P_6\}$, such that
\begin{itemize} 
\item $P_1 = \{1, 2,3,4\}$, 
\item $P_2 =\{5,6,7\}$, 
\item $P_3 = \{8,9,10,11,12\}$, 
\item $P_4=\{13,14,15\}$, 
\item $P_5=\{16,17,18,19,20,21,22\}$, and
\item $P_6=\{23,24,25,26,27\}$.
\end{itemize}
Again, we firstly evaluate the performance of our sampling proposal in the $RA$-random arrival rule estimation. As in the previous example, the task of determining the adequate sampling size is open and a bound of the absolute error is required to face this problem. Thus, we assume as bearable $\varepsilon=10$ euros.  Table \ref{sampling_sizes_ex2} provides the required sampling sizes that ensure an absolute error smaller or equal than 10 with probability at least $1-\alpha$. Notice that these values coincide with those ones used in the random arrival rule estimation in \cite{saavedrasaavedra}.

\begin{table}[ht]
	\begin{center}
		\resizebox{0.9\textwidth}{!}{	
			 \begin{tabular}{|p{0.012\textwidth}||p{0.06\textwidth}|p{0.1\textwidth}|p{0.1\textwidth}|||p{0.018\textwidth}||p{0.06\textwidth}|p{0.1\textwidth}|p{0.1\textwidth}|||p{0.015\textwidth}||p{0.06\textwidth}|p{0.1\textwidth}|p{0.1\textwidth}|}\hline
				$i$  & Union & $c_i$ & $RA^{RA}_i$ & $i$ & Union & $c_i$ & $RA^{RA}_i$ & $i$ & Union & $c_i$ & $RA^{RA}_i$  \tabularnewline\hline
1&1&15720.66&3930.1650&10&3&126857.13&24329.537&19&5&248008.45&31707.54\\
2&1&25532.20&6383.0500&11&3&248338.50&48604.527&20&5&169534.83&21920.58\\
3&1&32960.44&8240.1100&12&3&227091.64&44263.446&21&5&240404.84&30784.55\\
4&1&13664.61&3416.1525&13&4&63069.72&15767.430&22&5&250845.44&32051.92\\
5&2&8173.76&2043.4400&14&4&15915.98&3978.995&23&6&70752.96&16249.86\\
6&2&3904.17&976.0425&15&4&10059.72&2514.930&24&6&140679.05&32955.27\\
7&2&14869.04&3717.2600&16&5&530070.44&61413.004&25&6&227684.44&53920.80\\
8&3&289753.13&57070.6592&17&5&121229.15&15630.714&26&6&234125.14&55710.57\\
9&3&250962.13&49147.4723&18&5&233163.45&29907.336&27&6&264726.58&64579.14\\\hline
		\end{tabular}}
		\caption{Unions, claims and the $RA$-random arrival rule for the bankruptcy problem with a priori unions $(N,c,E,P)$.}\label{claims_ra_ex2}
	\end{center}
\end{table}

\vspace*{0.2 cm}\begin{table}[ht]
	\begin{center}
		\resizebox{0.93\textwidth}{!}{	
			 \begin{tabular}{|p{0.1\textwidth}|p{0.11\textwidth}|p{0.11\textwidth}|p{0.11\textwidth}|p{0.11\textwidth}|p{0.11\textwidth}|p{0.11\textwidth}|p{0.11\textwidth}|p{0.11\textwidth}|p{0.11\textwidth}|}\hline
				$i$&1&2&3&4&5&6&7&8&9\tabularnewline\hline
				$\alpha=0.1$&3701814&9764489&16272677&2796840&1000730&228313&3311608&1257561621&943385911\tabularnewline
				$\alpha=0.05$&4558333&12023778&20037820&3443967&1232277&281140&4077842&1548533981&1161664858\tabularnewline
				$\alpha=0.01$&6547109&17269687&28780212&4946551&1769913&403800&5856982&2224150880&1668492876\tabularnewline\hline\hline
				$i$&10&11&12&13&14&15&16&17&18\tabularnewline\hline
				$\alpha=0.1$&241047575&923764163&772458749&59581964&3794371&1515811&4208624456&220133999&814317838\tabularnewline
				$\alpha=0.05$&296820732&1137503064&951188873&73367932&4672306&1866536&5182408462&271068211&1002733244\tabularnewline
				$\alpha=0.01$&426321993&1633789424&1366187372&105377959&6710807&2680895&7443464905&389333787&1440220269\tabularnewline\hline\hline
				$i$&19&20&21&22&23&24&25&26&27\tabularnewline\hline
				$\alpha=0.1$&921310369&430517563&865684052&942508822&74982900&296436623&776496865&821049050&1049707018\tabularnewline
				$\alpha=0.05$&1134481516&530129947&1065984481&1160584829&92332309&365025599&956161322&1011021912&1292586352\tabularnewline
				$\alpha=0.01$&1629449590&761422741&1531067674&1666941636&132616390&524284269&1373329272&1452125238&1856534702\tabularnewline\hline
		\end{tabular}}
		\caption{Sampling sizes for estimating the $RA$-random arrival rule with $w_i=c_i$ for all $i\in N$.}\label{sampling_sizes_ex2}
	\end{center}
\end{table}

Table \ref{est_bankruptcy_ex2}  shows the comparative study of estimating the $RA$-random arrival rule with sampling sizes equal to  $\ell=10^3$, $\ell=10^5$ and $\ell=10^7$. We obtain the absolute error and the estimated variances. Both measures reduce when sampling size enlarges.

\begin{table}[ht]
	\begin{center}
		\resizebox{0.875\textwidth}{!}{	
			 \begin{tabular}{|p{0.02\textwidth}||p{0.115\textwidth}|p{0.11\textwidth}|p{0.14\textwidth}||p{0.115\textwidth}|p{0.095\textwidth}|p{0.14\textwidth}||p{0.115\textwidth}|p{0.095\textwidth}|p{0.14\textwidth}|}\hline
			 \multirow{2}{*}{$i$}& \multicolumn{3}{c||}{$\ell=10^3$}&\multicolumn{3}{c||}{$\ell=10^5$}&\multicolumn{3}{c|}{$\ell=10^7$}\tabularnewline\cline{2-10}
				&  Abs. error & Err. th., $\alpha=0.01$ & Estimated variance & Abs. error & Err. th., $\alpha=0.01$ & Estimated variance & Abs. error & Err. th., $\alpha=0.01$  & Estimated variance \tabularnewline\hline
  1 & 125.7650 & 809.1420 & 45334.220 & 30.8120 & 80.9142 & 460.9545 & 2.3460 & 8.0914 & 4.6357\\
2 & 204.2580 & 1314.1418 & 119580.700 & 50.0430 & 131.4142 & 1215.8860 & 3.8090 & 13.1414 & 12.2279\\
3 & 263.6840 & 1696.4731 & 199283.100 & 64.6020 & 169.6473 & 2026.2940 & 4.9180 & 16.9647 & 20.3779\\
4 & 109.3165 & 703.3172 & 34251.460 & 26.7825 & 70.3317 & 348.2659 & 2.0385 & 7.0332 & 3.5024\\
5 & 98.0850 & 420.7033 & 12116.460 & 1.9620 & 42.0703 & 125.1892 & 0.1980 & 4.2070 & 1.2528\\
6 & 46.8500 & 200.9475 & 2764.327 & 0.9370 & 20.0948 & 28.5615 & 0.0945 & 2.0095 & 0.2858\\
7 & 178.4280 & 765.3092 & 40095.700 & 3.5690 & 76.5309 & 414.2752 & 0.3600 & 7.6531 & 4.1457\\
8 & 2389.0008 & 14913.5874 & 12219697.000 & 142.2692 & 1491.3587 & 119778.8000 & 51.6008 & 149.1359 & 1198.2610\\
9 & 41.3277 & 12917.0154 & 9181235.000 & 481.0923 & 1291.7015 & 90231.8900 & 26.4477 & 129.1702 & 909.3835\\
10 & 601.4630 & 6529.3337 & 2499025.000 & 195.9270 & 652.9334 & 24345.2700 & 3.7930 & 65.2933 & 244.9093\\
11 & 3954.4333 & 12781.9772 & 9374676.000 & 236.1267 & 1278.1977 & 88777.9600 & 75.8433 & 127.8198 & 891.6024\\
12 & 882.3836 & 11688.4018 & 7586319.000 & 613.1464 & 1168.8402 & 73889.4900 & 19.8764 & 116.8840 & 747.5239\\
13 & 441.4900 & 3246.1971 & 731718.400 & 6.9400 & 324.6197 & 7456.1670 & 5.6100 & 32.4620 & 74.5658\\
14 & 111.4120 & 819.1951 & 46598.180 & 1.7510 & 81.9195 & 474.8327 & 1.4170 & 8.1920 & 4.7486\\
15 & 70.4180 & 517.7736 & 18615.470 & 1.1070 & 51.7774 & 189.6905 & 0.8950 & 5.1777 & 1.8970\\
16 & 7793.6461 & 27282.7141 & 25836109.000 & 937.2361 & 2728.2714 & 235204.1000 & 29.0139 & 272.8271 & 2317.5760\\
17 & 27.5657 & 6239.6617 & 1613480.000 & 18.1543 & 623.9662 & 16016.9000 & 3.2543 & 62.3966 & 160.2384\\
18 & 4407.6956 & 12000.9178 & 4981154.000 & 349.8544 & 1200.0918 & 57667.6800 & 33.7644 & 120.0092 & 571.7484\\
19 & 3508.1921 & 12764.9896 & 5804428.000 & 166.3121 & 1276.4990 & 63854.2000 & 20.0379 & 127.6499 & 641.8360\\
20 & 2743.8708 & 8725.9540 & 2782531.000 & 73.6392 & 872.5954 & 31100.1700 & 20.5992 & 87.2595 & 310.6117\\
21 & 2543.1383 & 12373.6319 & 5517885.000 & 11.6317 & 1237.3632 & 60537.6300 & 8.8717 & 123.7363 & 605.0563\\
22 & 1249.4167 & 12911.0094 & 6371426.000 & 271.0833 & 1291.1009 & 65942.4400 & 5.2133 & 129.1101 & 655.3291\\
23 & 270.7066 & 3641.6533 & 866177.300 & 132.7734 & 364.1653 & 8816.2240 & 15.1366 & 36.4165 & 87.5347\\
24 & 526.4682 & 7240.7477 & 3420345.000 & 14.4982 & 724.0748 & 34494.1300 & 37.9682 & 72.4075 & 344.9935\\
25 & 1666.5565 & 11718.9132 & 8941583.000 & 73.9065 & 1171.8913 & 87685.0700 & 23.7835 & 117.1891 & 876.5092\\
26 & 356.3484 & 12050.4159 & 9361653.000 & 170.7716 & 1205.0416 & 92290.5300 & 55.8316 & 120.5042 & 924.1232\\
27 & 813.5110 & 13625.4714 & 11504327.000 & 376.6790 & 1362.5471 & 117932.0000 & 67.1510 & 136.2547 & 1174.0490\\\hline
		\end{tabular}}
		\caption{Estimation of the $RA$-random arrival rule for $(N,c,E,P)$ for several values of $\ell$.}\label{est_bankruptcy_ex2}
	\end{center}
\end{table}

The results on Table \ref{est_bankruptcy_ex2} justifies the usage of this methodology for estimating the $RA$-random arrival rule because of its correct performance also in this example. However, the theoretical values of the error provided by  Corollary \ref{saavedra_cor2} are again larger than  the observed values. We conclude with a small simulation to check it. Table \ref{100_simus_ex2} displays the minimum, maximum and  averaged  absolute errors for $1000$ estimations of the random arrival rule by using $\ell=10^4$ in this example.

\begin{table}[h!]
	\begin{center}
		\resizebox{0.9\textwidth}{!}{	
			 \begin{tabular}{|p{0.22\textwidth}|p{0.09\textwidth}|p{0.09\textwidth}|p{0.095\textwidth}|p{0.095\textwidth}|p{0.095\textwidth}|p{0.095\textwidth}|p{0.095\textwidth}|p{0.095\textwidth}|p{0.095\textwidth}|}\hline
			$i$&1&2&3&4&5&6&7&8&9\\\hline
			Theoretical, $\alpha=0.1$&255.8732&415.5681&536.4719&222.4084&133.0380&63.5452&242.0120&4716.0904&4084.7189\\\hline
			Maximum&132.0535&214.4705&276.8677&114.7827&93.1809&44.5075&169.5071&2719.3987&3152.5928\\
			Average&50.1646&81.4733&105.1768&43.6038&27.4720&13.1219&49.9748&885.1569&737.8555\\
			Minimum&3.1441&5.1064&6.5921&2.7329&0.8174&0.3904&1.4869&8.9433&0.4221\\\hline\hline
			$i$&10&11&12&13&14&15&16&17&18\\\hline
			Theoretical, $\alpha=0.1$&2064.7566&4042.0161&3696.1972&1026.538&259.0523&163.7344&8627.5517&1973.1543&3795.0234\\\hline
			Maximum&1529.7863&2117.6095&2602.0905&1034.343&261.0221&164.9794&3881.3046&1010.1539&1935.7362\\
			Average&421.6090&707.1321&682.5632&207.184&52.2840&33.0462&1183.1003&308.9907&591.7395\\
			Minimum&2.7277&9.6001&9.3514&0.000&0.0000&0.0000&19.7829&12.1751&13.4823\\\hline\hline
			$i$&19&20&21&22&23&24&25&26&27\\\hline
			Theoretical, $\alpha=0.1$&4036.6441&2759.3890&3912.8860&4082.8197&1151.5919&2289.7255&3705.8457&3810.6761&4308.7524\\\hline
			Maximum&1680.3395&1308.7188&2253.2597&1796.4386&1002.4567&1706.3678&2966.9821&2649.1133&3167.3045\\
			Average&614.0042&425.7196&583.4490&527.3767&270.0250&482.3436&744.9507&724.6656&860.9238\\
			Minimum&8.8083&7.5728&9.0623&4.6156&2.6795&8.5164&9.7456&2.9140&25.9181\\\hline
		\end{tabular}}
		\caption{Summary of the absolute errors in the $1000$ simulations.}\label{100_simus_ex2}
	\end{center}
\end{table}

\newpage
{
\section{An application: new milk quotas for Galicia based on its territorial organization}\label{sec5}

The sampling procedure described along this work will be applied on the real bankruptcy situation which arises from the end of the milk quotas regime imposed by the European Union (EU) to regulate the milk market. Notice that the analysis of this class of situations can be modeled as a bankruptcy problem when the maximum of tons of milk in 2014-2015 imposed for Galicia reduces by $\rho\%$ of the total, with $\rho\in (0,100]$. 
Figure \ref{mapa_gal_quotas} in Appendix A shows represents graphically the milk quota of each council in this period taken as reference. From Table \ref{claims_190concellos} in Appendix A, the set of involved agents is given by the 190  councils of Galicia indicated there, that is, $N=\{1,\dots,190\}$. Thus, the elements of the bankruptcy problem are formally defined as follows. The resources to be allocated among the different councils (the estate) corresponds to the $(100 -\rho)\%$ of the aggregate milk quota for the region in 2015 that is, of $2229811.281$ tons of milk. We take as claims of the councils the maximum bound of milk (in tons) given by the individual quotas in the period 2014-2015. These amounts describe the maximum milk production capacities of each municipality.  The random arrival rule for bankruptcy problems considering the provinces and also the counties as a structure of a priori unions will provide a new milk quota for each council under the fairness criteria that bases the Owen value for TU-games. The main difficulty for its obtaining is that the large set of players that are involved, implicitly increases the computational complexity for the exact computation. For this reason, we approximate the $RA-$random arrival rule of each council in Galicia by using simple random sampling with replacement described in Section \ref{samplingalgorithm} with $\ell=10^7$. In this illustration, we focus on the case when $\rho= 40$ for simplicity but, of course, similar discussions could be considered for other values of $\rho$. As consequence, the portion of the regional milk quota for 2015 we manage as estate is $60\%$ of the total. Table \ref{RA_190concellos_provincias} in Appendix B contains the estimations of the $RA-$random arrival rule when $\rho= 40$, $\rho= 10$ and $\rho= 5$. Estimations when counties are selected as systems of a priori unions are shown in Table \ref{RA_190concellos_counties} also in Appendix B. Therefore, they contain a total of six new system of quotas for the $190$ councils in Galicia when the four provinces or counties are considered as a priori unions.

Results obtained for $\rho=40$ were compared to the obtained ones in \cite{saavedrasaavedra} where  a priori unions were not be taken into account in order to estimate the allocation rule. The consideration of the provinces as a priori unions has different effects on councils depending on the province that they belong. Specifically, all councils in A Coruña reduce their milk quotas between a 5\% and a 7\% and all councils in Lugo increase their production between a 2\% and a 3\%.  However, the biggest effects are detected for councils of Pontevedra and Ourense. Concretely, councils colored in blue in Table \ref{RA_190concellos_provincias} increase a 11\% their milk quotas when provinces are established as a priori unions structure. Moreover, it can be easily checked that the production of Lalín and Silleda that are two of the three councils where the number of farms is largest increase considerably. Table \ref{sumary1} in Appendix A shows the percentages by provinces of the aggregated of the milk quotas in 2014-2015. Remark that provinces of Pontevedra and Ourense only represent around a $12.5\%$ of the total Galician production (see also Figure \ref{map11}). Therefore, the estimated $RA-$random arrival rule increases the allocated quotas of the councils that are located in the provinces with smaller aggregated quotas.

As for the effect of counties, all councils suffer variations on their milk quotas if a priori unions are considered. Specifically, the $RA-$random arrival rule causes decreasings smaller than a 1\% of the production in most of counties. Only councils represented in blue color in Table \ref{RA_190concellos_counties} have an increasing (slightly over 1\%) of their milk quotas. Therefore, all councils in the counties of Ordes, Lugo, Terra Chá, O Deza would increase they milk quotas. According Table \ref{sumary2}, they correspond to the biggest milk producers in the period 2014-2015. In this case, the $RA-$random arrival rule promotes the development of those counties with largest productions decreasing the milk quotas of the rest of counties in a similar proportion even when levels of production are specially low.

\subsection{Computation of milk quota systems using alternative bankruptcy rules with a priori unions}

The previous section is mainly devoted to the estimation of the $RA-$random arrival rule for the bankruptcy problem that arises in the milk conflict of Galicia. This rule that underlies this work is not the only one applicable in these settings. As we have previously mentioned, there are multiple extensions of classical rules in the bankruptcy literature for a priori unions context and that are based on different criteria. Here, we consider the extension of the Talmud rule, $\overline{T}(N,c,E,P)$, and the extension of the proportional rule, $\overline{P}(N,c,E,P)$, as in Section \ref{sec2} for the associated bankruptcy situation with a priori unions $(N,c,E,P)\in BU^N$, following \cite{borm2005constrained}.

Tables \ref{rules_190concellos} and \ref{rules_190concellos_counties} show the results obtained for the extended Talmud and the extended proportional rules when provinces and counties are considered as a priori unions. They were also treated in \cite{saavedrasaavedra} when  $\rho=40$ but there, a priori unions were not taken into account. As before, the role of provinces and counties will be discussed on the milk quotas that this extension of the Talmud rule provides in both cases. The approach of the proportional rule does not cause any change on the results in \cite{saavedrasaavedra} when provinces or counties are incorporated as a priori unions. Therefore, discussion is only focused on the case of the Talmud rule.

The effect of provinces cause some variations in the resulting milk quotas for extension of the Talmud rule to those settings of a priori unions. However, most of councils do not suffer any variation with respect to results in \cite{saavedrasaavedra}. Note that the only differences detected are not usually bigger than a 7\%. Specifically, the above-mentioned extension of the Talmud rule causes that Negreira, Ordes and Pastoriza increase their milk quotas between a 2\% and a 3\%; Curtis and Castro de Rei, a 4\%; Mazaricos, Santa Comba, Paradela, Sarria and Cospeito, between a 5\% and a 6\% and Arzúa and Frades, a 7\%. Note that many of them are the biggest producers of milk in their counties or, of course, they have remarkable levels of production. Additionally, results for some particular councils must be discussed. Only the councils of Lalín, Rodeiro and Silleda, all of them in the province of Pontevedra, decrease their allocations when provinces are considered as a priori unions. Concretely, the corresponding milk quotas decrease a  respectively. But councils of Mesía, Tordoia, Trazo, Taboada, Castro Verde, Guntín, Lugo, Pol and Guitiriz increase their milk quotas from 8\% to 12\% when provinces are involved in the computation of the rule. Hence, it seems that those councils with the highest production in a province with a low milk quota such as Pontevedra (see Figure \ref{map11}) are penalyzed in favour of those councils with the largest levels of production belonging to those provinces with the highest milk quotas.

\begin{table}[h!]\begin{center}
		\resizebox{\textwidth}{!}{	\begin{tabular}{|lcc||lcc||lcc|}
				\hline
				County&  Milk quota& $\%$ & County & Milk quota&$\%$& County&Milk quota&   $\%$\\
				\hline
			 A Barcala&22902.45&1.712&Terra de Melide&25701.37&1.921 &Baixa Limia&33.84&0.003\\
			 A Coruña&6287.29&0.470 &Terra de Soneira&20107.70&1.503 &A Limia&2466.17&0.184\\
			 Arzúa&51642.84&3.860 &Xallas&63755.62&4.765 &Allariz-Maceda&3182.62&0.238\\
			 Bergantiños&30796.28&2.302 &A Fonsagrada&5852.98&0.437 &O Carballiño&1057.37&0.079\\
			 Betanzos&42513.86&3.178 &A Mariña Central&8320.56&0.622&Ourense&221.07&0.017\\
			 Eume&16831.24&1.258&A Mariña Occidental&11.32&0.001 &Terra de Caldelas&162.46&0.012\\
			 Ferrol&7459.15&0.558&A Mariña Oriental&33833.70&2.529 &Terra de Celanova&501.92&0.038\\
			 Fisterra&16757.41&1.253&A Ulloa&24834.73&1.856 &Terra de Trives&203.72&0.015\\
			 Muros&19.30&0.001 &Chantada&59458.77&4.444  &Verín&559.91&0.042\\
			 Noia&3405.23&0.255&Lugo&142791.67&10.673  &Viana&1279.78&0.096\\
			 O Barbanza&133.10&0.010  &Meira&32100.85&2.399 &Caldas&657.54&0.049\\
			 Ordes&147956.25&11.059 &Os Ancares&3601.96&0.269&O Deza&159395.01&11.914\\
			 Ortegal&2299.32&0.172&Sarria&69924.08&5.226&O Salnés&98.23&0.007\\
			 Santiago&34508.55&2.579&Terra Chá&248631.03&18.584&Pontevedra&298.72&0.022\\
			 Sar&7366.87&0.551&Terra de Lemos&21959.60&1.641&Tabeirós-T.de Montes&16003.36&1.196\\

				\hline
	\end{tabular}}\end{center}\caption{Percentages by counties of the aggregated milk quotas computed using the extended Talmud rule with counties as a priori unions.}\label{sumarynova}\end{table}

As regards the role of counties, milk quotas vary considerably when they are included as a priori unions. For instance, milk quotas of Arzúa, Mazaricos Santa Comba, Chantada, Taboada, Pol and Sarria decrease around a 29\%, 34\%, 35\%, 31\%, 20\%, 22\% and 23\%, respectively. However, other councils will increase the tons of milk to produce notably. In particular, Frades (13\%), Mesía (16\%), Ordes (18\%), Tordoia (20\%), Trazo (21\%), Castro Verde (26\%), Friol (27\%), Guntín (20\%), Lugo (18\%), Portomarín (20\%), Castro de Rei (11\%), Cospeito (15\%), Guitiriz (28\%) and Vilalba (24\%). In this case, the effect of the counties seems to penalize the milk quotas of councils that are big producers in favour of more intermediate producers. Table \ref{sumarynova} shows the percentages of production for these new milk quotas by counties. The comparison of these results with the contained in Table \ref{sumary2} in Appendix A allows to extract a similar conclusion. Remark that this new milk quotas tends to balance the percentages between counties again penalize the biggest producers. According results obtained, the extended version of Talmud rule is the allocation method where counties produce the most extreme changes in the resulting system of quotas.

\section{Concluding remarks}\label{sec6}

In this work, we have described a computational procedure to estimate the $RA-$random arrival rule incorporating a priori unions for bankruptcy problems. This proposal is based on simple random sampling with replacement following the scheme in \cite{saavedrasaavedra}. The determination of this rule is specially complicated when bankruptcy situations involved a large set of agents. Moreover, this new approach can be computed in parallel reducing the required time and the complexity with respect to the exact computation of the $RA-$random arrival rule for bankruptcy problems.

Some theoretical results to ensure that our proposal correctly approximates the real value were also provided following \cite{Saavedra2018a}. In particular, this allows to  determine the adequate sampling size for estimations. The performance of our sampling method has been also checked on two well-known examples in the literature.

Finally, it is worth mentioning the application of this proposal of sampling on a practical case. Concretely, the milk problem arisen in Galicia after the suppression of the European milk quotas in April 2015 and that led to a social conflict in the region is also treated here. In this case, the role of the administrative territorial divisions of Galicia in provinces and counties to solve this problem is checked. Concretely, the $RA-$random arrival rule in bankruptcy taking the provinces and counties as a priori unions is introduced as a mechanism of determining a new system of milk quotas based on a fairness criteria as solution. Besides, we compare these two resulting system of quotas with the ones obtained under the corresponding extensions of classical rules in bankruptcy literature, as the Talmud and proportional rules as well as the milk quotas established in \cite{saavedrasaavedra} where a priori unions were not considered. The role of provinces and counties as a priori unions is not insignificant at all in order to establish new milk quotas depending on the objectives.

For instance, the effect of considering the organization in provinces as a priori unions system in the estimation of the $RA-$random arrival rule increases the allocated quotas of those councils that are located in the provinces with smaller aggregated quotas. However, when counties are taken as a priori unions, milk quotas of those ones that present largest levels of production increase slightly their milk quotas. Moreover, the rest of counties decrease their levels of production in a similar proportion even when they are not very representative.

As for Talmud rule under the existence of a priori unions structure, the role of counties penalyze the milk quotas of councils that are big producers in favour of intermediate ones. If provinces are established as a priori unions, councils that have largest production in a province with low milk quota suffer production reductions in favour of councils that present the largest production in provinces with the highest milk quotas. 

Therefore, territorial organization of Galicia in provinces and counties could play an decisive role in order to establish new system of quotas attending to the specific needs of public administrations and producers. For instance, those rules that tend to balance the milk quotas between councils could clearly contribute to structure and revitalize Galician territory and those others ones that enhance the areas that are already big milk producers would specialize the dairy sector.

\section*{Acknowledgments}
A. Saavedra-Nieves acknowledges the financial support of \textit{Ministerio de Econom\'{\i}a y Competitividad} of the Spanish government under grant MTM2017-87197-C3-2-P. P. Saavedra-Nieves acknowledges the financial support of \textit{Ministerio de Econom\'{\i}a y Competitividad} of the Spanish government under grants MTM2016-76969P and MTM2017-089422-P.

\renewcommand{\BBAA}{and}
\bibliographystyle{apacite}
\bibliography{referencias_alex}

\clearpage\section*{Appendix A. Milk quotas for $190$ councils in Galicia in the \mbox{period} 2014-2015}

Table \ref{claims_190concellos} shows the milk quotas for $190$ councils in Galicia in the period 2014-2015. Figure \ref{mapa_gal_quotas} shows the councils that are milk producers in gray color and through the size of a corresponding bullet indicates the proportion of its milk quota. These data can be obtained from the website of Consellería de Medio Rural of Xunta de Galicia\footnote{Galician Government website: \url{https://mediorural.xunta.gal/}}. Moreover, we also show the number of farms in each council the councils distribution by provinces and counties, respectively.

A brief summary of information in Table \ref{claims_190concellos} is shown in Tables \ref{sumary1} and \ref{sumary2}. Concretely, Table \ref{sumary1} contains the aggregated milk quotas in the period 2014-2015 by provinces as well as the corresponding percentages. Table \ref{sumary2} also shows the the aggregated milk quotas in the period 2014-2015 but, in this case, by counties. Moreover, the seven counties with a production percentage bigger than the 5\% are represented in blue color. Remark that they represent practically the 60\% of the milk quotas in Galicia. Counties that are between the 3\% and the 5\% were represented in gray color. They are Arzúa, Betanzos, Santiago and A Mariña Oriental.

\begin{table}[h!]\begin{center}
			\begin{tabular}{|ccc|}
				\hline
				Province& Milk quotas& $\%$\\
								\hline
A Coruña&	917478.555&	41.146 \\
Lugo&	1034937.150&	46.414 \\
Ourense&	19337.713&	0.867 \\
Pontevedra	&258057.863&	11.573 \\				\hline
\end{tabular}\end{center}\caption{Aggregated milk quotas in the period 2014-2015 by provinces and percentages of the total.}\label{sumary1}\end{table}

\begin{table}[h!]\begin{center}
		\resizebox{\textwidth}{!}{	\begin{tabular}{|lcc||lcc||lcc|}
			\hline
		County& Milk quota& $\%$ & County& Milk quota& $\%$& County& Milk quota& $\%$\\
			\hline
			A Barcala&45804.891&2.053&	Terra de Melide&51402.731&2.304	&	Baixa Limia&67.674&0.003\\
			A Coruña&12574.571&0.564&	Terra de Soneira&40215.401&1.802	&	A Limia&4932.333&0.221\\
		\cellcolor{GG}	Arzúa&103285.678&4.629&	\cellcolor{GGG}Xallas&127511.247&5.715	&	Allariz-Maceda&6365.234&0.285\\
			Bergantiños&61592.554&2.761&	A Fonsagrada&11705.957&0.525&	O Carballiño&2114.747&0.095\\
			\cellcolor{GG}Betanzos&85027.716&3.811&A Mariña Central&16641.110&0.746&	Ourense&442.136&0.020\\
		Eume&33662.483&1.509&		A Mariña Occidental&22.638&0.001	&	Terra de Caldelas&1736.200&0.078\\
			Ferrol&14918.297&0.669&			\cellcolor{GG}A Mariña Oriental&67667.409&3.033	&Terra de Celanova	&1003.846&0.045\\
			Fisterra&33514.821&1.502&	A Ulloa&49669.460&2.226&Terra de Trives&407.444&0.018\\
			Muros&38.603&0.002&	\cellcolor{GGG}Chantada&118917.530&5.330		&	Verín&1119.825&0.050\\
			Noia&6810.455&0.305	&		\cellcolor{GGG}Lugo&207338.828&9.293&	Viana&2559.564&0.115\\
		O Barbanza&266.209&0.012	&	Meira&64201.693&2.877	&	Caldas&1315.070&0.059\\
			\cellcolor{GGG}Ordes&212503.407&9.524&	Os Ancares&7203.916&0.323	&		\cellcolor{GGG}O Deza&223942.163&10.037\\
		Ortegal&4598.637&0.206	&	\cellcolor{GGG}Sarria&134471.233&6.027	&	O Salnés&196.463&0.009\\
		\cellcolor{GG}	Santiago&69017.109&3.093&		\cellcolor{GGG}Terra Chá&313178.182&14.036		&	Pontevedra&597.445&0.027\\
			Sar&14733.745&0.660&Terra de Lemos&43919.194&1.968	&	Tabeirós-T. de Montes&32006.722&1.435\\
		 	\hline
	\end{tabular}}\end{center}\caption{Aggregated milk quotas in the period 2014-2015 by counties and percentages of the total.}\label{sumary2}\end{table}

\begin{landscape}
{\begin{table}[h!]
	\begin{center}
		\resizebox{1.6\textheight}{0.47\textwidth}{	
			 \begin{tabular}{|p{0.115\textwidth}|p{0.118\textwidth}|p{0.142\textwidth}|p{0.065\textwidth}|p{0.096\textwidth}||p{0.054\textwidth}|p{0.17\textwidth}|p{0.16\textwidth}|p{0.065\textwidth}|p{0.099\textwidth}||p{0.09\textwidth}|p{0.17\textwidth}|p{0.21\textwidth}|p{0.065\textwidth}|p{0.098\textwidth}||p{0.12\textwidth}|p{0.21\textwidth}|p{0.19\textwidth}|p{0.065\textwidth}|p{0.115\textwidth}|}\hline
				Prov.& County & Council& Farms&  $c_i$&Prov.& County & Council& Farms& $c_i$ &Prov. & County & Council& Farms&  $c_i$&Prov.& County & Council& Farms& $c_i$\\\hline
A Coruña & A Barcala & A Baña &  105    & 19253.147 &  &  & Muxía &  86    & 16694.545 &  &  & Palas de Rei &  117    & 26541.407 &  &  & Maceda &  6    & 3685.430\\\cdashline{7-10}\cdashline{12-15}
 &  & Negreira &  140    & 26551.744 &  & Muros & Carnota &  1    & 38.603 &  & Chantada & Carballedo &  148    & 23757.953 &  &  & Xunqueira de A. &  3    & 864.995\\\cdashline{2-5}\cdashline{7-10}
 & A Coruña & A Coruña &  1    & 606.673 &  & Noia & Lousame &  23    & 4918.423 &  &  & Chantada &  311    & 54438.744 &  &  & Xunqueira de E.&  1    & 71.678\\\cdashline{17-20}
 &  & Abegondo &  27    & 2957.506 &  &  & Noia &  1    & 95.521 &  &  & Taboada &  212    & 40720.833 &  & O Carballiño & O Carballiño &  1    & 64.606\\\cdashline{12-15}
 &  & Arteixo &  15    & 3084.636 &  &  & Outes &  21    & 1796.511 &  & Lugo & Castroverde &  173    & 36769.831 &  &  & O Irixo &  9    & 286.276\\\cdashline{7-10}
 &  & Bergondo &  3    & 360.532 &  & O Barbanza & Rianxo &  1    & 21.320 & & & Friol &  165    & 27096.989 &  &  & Piñor &  8    & 1054.850\\
 &  & Cambre &  16    & 2313.154 &  &  & Ribeira &  4    & 244.889 &  &  & Guntín &  174    & 42735.528 &  &  & San Amaro &  3    & 156.146\\\cdashline{7-10}
 &  & Carral &  17    & 2201.589 &  & Ordes & Cerceda &  70    & 11547.136 &  & & Lugo &  233    & 45298.976 &  &  & S. Crist. de Cea &  9    & 552.869\\\cdashline{17-20}
 &   & Culleredo &  7    & 869.505 &  &  & Frades &  183    & 48950.157 &  & & O Corgo &  81    & 16027.173 &  & Ourense & Barbadás &  1    & 52.630\\
 &  & Oleiros &  1    & 180.976 &  &  & Mesía &  198    & 43066.524 &  & & Outeiro de Rei &  82    & 14598.522 &  &  & Coles &  1    & 208.133\\\cdashline{2-5}
 & Arzúa & Arzúa &  250    & 50811.051 &  &  & Ordes &  170    & 26674.213 &  & & Portomarín &  98    & 24811.809 &  &  & Taboadela &  2    & 181.373\\\cdashline{12-15}\cdashline{17-20}
 &  & Boimorto &  102    & 17949.237 &  &  & Oroso &  55    & 8543.421 &  & Meira & Meira &  56    & 11425.526 &  & Terra de Caldelas & A Teixeira &  1    & 324.910\\\cdashline{17-20}
 &  & O Pino &  78    & 11176.461 &  &  & Tordoia &  202    & 37560.049 & & & Pol &  141    & 42774.089 &  & T. de Celanova & A Bola &  2    & 13.253\\
 &  & Touro &  111    & 23348.929 &  &  & Trazo &  136    & 36161.907 &  &  & Ribeira de Piquín &  19    & 1762.278 &  &  & A Merca &  1    & 31.593\\\cdashline{2-5}\cdashline{7-10}
 & Bergantiños & A Laracha &  105    & 13905.151 &  & Ortegal & Cerdido &  6    & 1165.265 &  &  & Ríotorto &  43    & 8239.800 &  &  & Cartelle &  11    & 464.957\\\cdashline{12-15}
 &  & Cab. de Berg. &  37    & 9509.237 &  &  & Mañón &  2    & 298.197 &  & Os Ancares & As Nogais &  4    & 268.871 &  &  & Celanoa &  7    & 442.096\\
  &  & Carballo &  73    & 10655.940 & &  & Ortigueira &  10    & 3135.175 &  & & Baralla &  32    & 5458.796 &  &  & Ramirás &  2    & 51.947\\\cdashline{7-10}\cdashline{17-20}
 &  & Coristanco &  65    & 10970.444 &  & Santiago & Ames &  16    & 1958.122 &  & & Becerreá &  15    & 1473.789 &  & Terra de Trives & Manzaneda &  1    & 118.854\\
 &  & Laxe &  7    & 592.689 & &  & Boqueixón &  61    & 13523.385 &  &  & Ped. do Cebreiro &  1    & 2.460 &  &  & Río &  5    & 288.590\\\cdashline{12-15}\cdashline{17-20}
 &  & Malpica &  15    & 2421.117 &  &  & Brión &  57    & 12969.246 &  & Sarria & Láncara &  85    & 16596.042 &  & Verín & Riós &  7    & 1119.825\\\cdashline{17-20}
 & & Ponteceso &  54    & 13537.976 &  &  & Santiago  &  60    & 12528.073 &  &  & O Incio &  31    & 4575.134 &  & Viana & A Gudiña &  4    & 264.956\\\cdashline{2-5}
 & Betanzos & Aranga &  94    & 18626.978 &  & & Teo &  14    & 1968.363 &  &  & O Páramo &  89    & 20147.930 &  &  & A Mezquita &  6    & 805.467\\
 &  & Betanzos &  2    & 199.583 &  &  & Val do Dubra &  117    & 22630.365 & &  & Paradela &  107    & 27541.073 &  &  & Viana do Bolo &  9    & 1337.802\\
 &  & Cesuras &  71    & 8784.328 & & & Vedra &  36    & 3439.555 &  &  & Samos &  58    & 10757.825 &  &  & Vilariño de Conso &  1    & 151.339\\\cdashline{7-10}\cdashline{16-20}
 &  & Coirós &  1    & 52.966 &  & Sar & Padrón &  3    & 1383.377 &  &  & Sarria &  234    & 53345.565 & Pontevedra & Caldas & Caldas de Reis &  1    & 5.938\\
 &  & Curtis &  122    & 27152.543 &  &  & Rois &  68    & 13350.368 &  &  & Triacastela &  6    & 1507.664 &  &  & Cuntis &  31    & 837.933\\\cdashline{7-10}\cdashline{12-15}
 &  & Irixoa &  63    & 8799.154 &  & Terra de Melide & Melide &  114    & 15579.976 &  & Terra Chá & A Pastoriza &  313    & 87103.413 &  &  & Moraña &  2    & 40.681\\
 &  & Miño &  8    & 1643.428 & &  & Santiso &  115    & 17539.229 &  &  & Abadín &  119    & 12328.510 &  &  & Portas &  1    & 89.872\\
 &  & Oza dos Ríos &  26    & 3032.885 &  &  & Sobrado &  91    & 11865.545 &  &  & Begonte &  47    & 6273.036 &  &  & Valga &  1    & 340.646\\\cdashline{17-20}
 &  & Paderne &  15    & 3391.447 &  &  & Toques &  66    & 6417.981 &  &  & Castro de Rei&  259    & 71736.134 &  & O Deza & Agolada &  100    & 10045.584\\\cdashline{7-10}
 &  & Vilarmaior &  21    & 2270.729 &  & T. de Soneira & Vimianzo &  129    & 22799.899 &  &  & Cospeito &  255    & 54336.024 &  &  & Dozón &  87    & 15730.244\\
 &  & Vilasantar &  65    & 11073.675 & & & Zas &  136    & 17415.502 &  &  & Guitiriz &  185    & 37429.143 &  &  & Lalín &  478    & 82356.544\\\cdashline{2-5}\cdashline{7-10}
 & Eume & A Capela &  56    & 9049.543 &  & Xallas & Mazaricos &  307    & 61046.064 &  & & Vilalba &  169    & 24364.386 &  &  & Rodeiro &  275    & 48166.205\\
 &  & As Pontes &  22    & 2212.419 &  &  & Santa Comba &  389    & 66465.183 &  & & Xermade &  80    & 19607.536 &  &  & Silleda &  361    & 52106.510\\\cdashline{6-10}\cdashline{12-15}
 &  & Cabanas &  7    & 779.033 & Lugo & A Fonsagrada & A Fonsagrada &  6    & 1413.767 &  & Terra de Lemos & A Pobra de Brollón &  62    & 7978.205 &  &  & Vila de Cruces &  147    & 15537.076\\\cdashline{17-20}
 &  & Monfero &  159    & 21125.185 &  &  & Baleira &  52    & 10292.190 &  &  & Bóveda &  27    & 4271.243 &  & O Salnés & Meis &  1    & 29.518\\\cdashline{7-10}
 &  & Pontedeume &  4    & 496.303 &  & A Mariña C. & Alfoz &  14    & 1217.884 &  &  & Monforte de Lemos &  34    & 5877.472 &  &  & O Grove &  1    & 26.793\\\cdashline{2-5}
 & Ferrol & As Somozas &  11    & 970.062 &  &  & Foz &  14    & 3736.900 &  &  & O Saviñao &  129    & 23605.507 &  &  & Pontecesures &  1    & 8.076\\
 &  & Fene &  4    & 363.962 &  & & Lourenzá &  33    & 4404.102 &  &  & Pantón &  12    & 1187.648 &  &  & Ribadumia &  2    & 132.076\\\cdashline{17-20}
 &  & Moeche &  13    & 2623.408 &  &  & Mondoñedo &  47    & 4655.363 &  &  & Sober &  4    & 999.119 &  & Pontevedra & Campo Lameiro &  2    & 50.603\\\cdashline{11-15}
 &  & Mugardos &  1    & 330.711 &  &  & O Valadouro &  11    & 2626.861 & Ourense & Baixa Limia & Lobios &  1    & 67.674 &  &  & Cotobade &  1    & 220.001\\\cdashline{7-10}\cdashline{12-15}
 &  & Narón &  8    & 1087.991 &  & A Mariña Occ. & Ourol &  2    & 22.638 &  & A Limia & Baltar &  2    & 131.653 &  &  & Poio &  1    & 19.424\\\cdashline{7-10}
 &  & Neda &  3    & 369.915 &  & A Mariña Or. & A Pontenova &  25    & 5118.890 &  &  & Rairiz de Veiga &  2    & 34.773 &  &  & Pontevedra &  1    & 307.417\\\cdashline{17-20}
 &  & Ferrol &  1    & 471.831 &  &  & Barreiros &  57    & 17735.873 &  &  & Sandiás &  1    & 195.067 &  & Tabeirós-T. Montes & A Estrada &  216    & 19471.968\\
 &  & S. Sadurniño &  35    & 7752.548 &  & & Ribadeo &  96    & 27561.883 &  &  & Sarreaus &  6    & 1542.098 &  &  & Cerdedo &  4    & 66.908\\
 &  & Valdoviño &  9    & 947.869 &  &  & Trabada &  55    & 17250.763 &  &  & Trasmiras &  1    & 180.270 &  & & Forcarei &  121    & 12467.846\\\cdashline{2-5}\cdashline{7-10}\cline{16-20}
 & Fisterra & Cee &  5    & 544.511 &  & A Ulloa & Antas de Ulla &  33    & 4740.799 &  & & Xinzo de Limia &  2    & 2848.472 & Total &  &  &  2.132    & 2229811.281\\\cdashline{12-15}
 &  & Dumbría &  77    & 16275.765 &  &  & Monterroso &  84    & 18387.254 &  & Allariz-Maceda & Allariz &  4    & 1743.131 &  &  &  &  & \\\hline
		\end{tabular}}
		\caption{Milk quotas for $190$ councils in Galicia in the period 2014-2015 organized by provinces and counties.}\label{claims_190concellos}
	\end{center}
\end{table}}

\end{landscape}

 \begin{figure}[h!]
	\centering
	\includegraphics[width=0.66\textwidth,height=9.9 cm]{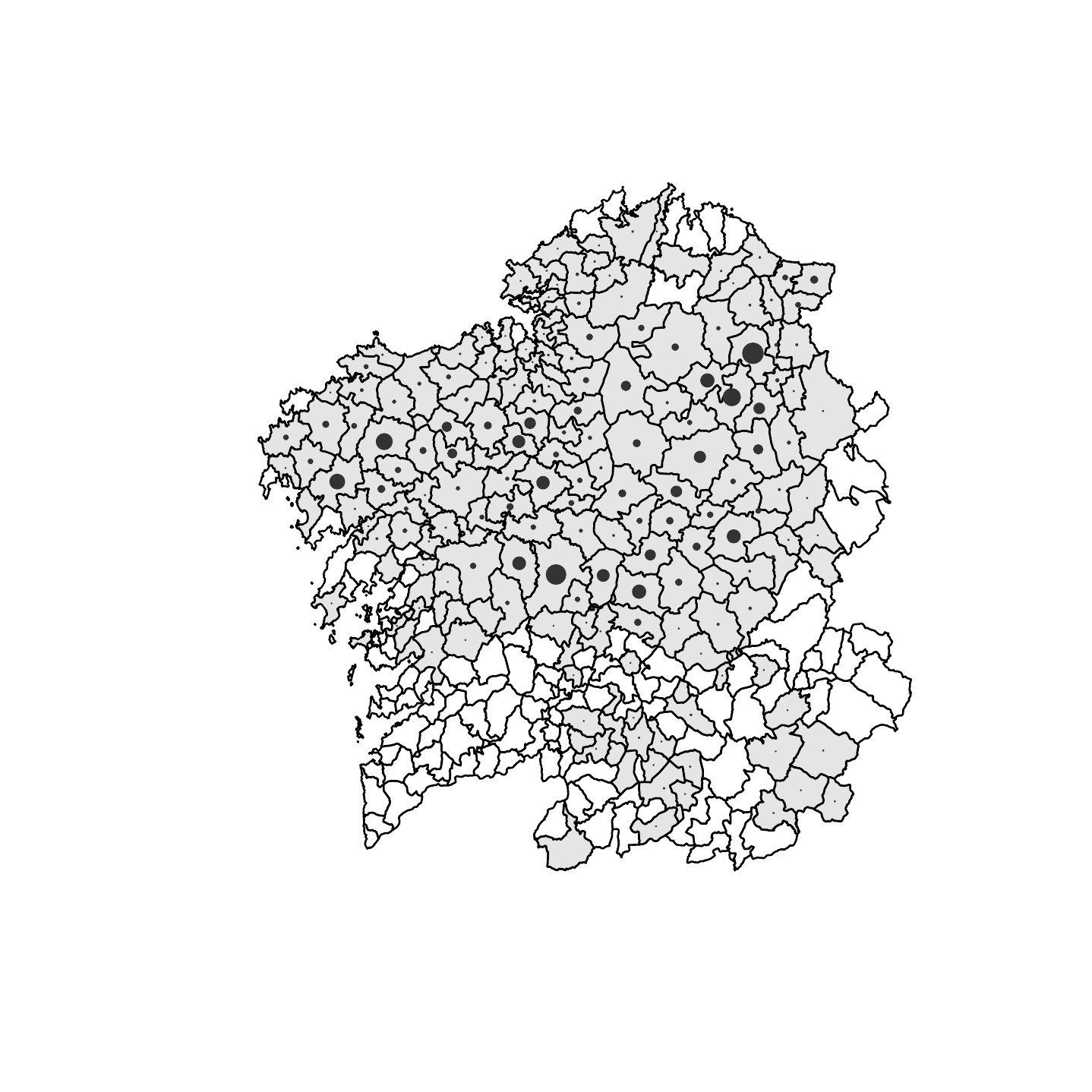}
	\caption{Map of Galicia and the involved councils (in gray). In dark gray, the size of the bullet of each council indicates the proportion of its milk quota in 2014-2015.}
	\label{mapa_gal_quotas}
\end{figure}

 \section*{Appendix B. New systems of milk quotas for $190$ councils in Galicia in the period 2014-2015.}

Table \ref{RA_190concellos_provincias} contains the new systems of milk quotas for the set of 190 councils in Galicia in the period 2014-2015 obtained from the estimations of the $RA-$rule when provinces are considered as a priori unions with $\ell=10^7$.  Table \ref{RA_190concellos_counties} shows the corresponding results when the counties are taken as a priori unions. Three values of the parameter $\rho$ were considered. Concretely, 
$\rho= 40$, $\rho= 10$ and $\rho= 5$. Therefore, the corresponding estates are $E=1337887$,  $E= 2006830$ and $E= 2118321$, respectively.

 Table \ref{rules_190concellos}  shows two new systems of milk quotas from Talmud and proportional rules for the set of 190 councils in Galicia in the period 2014-2015 when we impose a reduction of production equal to $40\%$ ($\rho=40$). The cases $\rho=10$ and $\rho=5$ can be supplied on request.
 
\begin{table}[h!]
	\begin{center}
		\resizebox{0.69\textwidth}{!}{	
			\begin{tabular}{|p{0.28\textwidth}|p{0.10\textwidth}|p{0.10\textwidth}|p{0.10\textwidth}||p{0.28\textwidth}|p{0.10\textwidth}|p{0.10\textwidth}|p{0.10\textwidth}|}\hline
				Council & $\rho=40$ & $\rho=10$ & $\rho=5$ & Council & $\rho=40$ & $\rho=10$ & $\rho=5$ \\\hline
				A Baña & 10905.670 & 17708.09 & 18479.1 & Monterroso & 11339.500 & 17060.45 & 17705.24\\
				Negreira & 14997.770 & 24437.15 & 25507.87 & Palas de Rei & 16340.110 & 24641.94 & 25584.47\\
				A Coruña & 345.908 & 556.932 & 580.745 & Carballedo & 14633.950 & 22050.84 & 22889.75\\
				Abegondo & 1685.140 & 2716.004 & 2832.416 & Chantada & 33314.350 & 50706.15 & 52679.46\\
				Arteixo & 1757.566 & 2832.658 & 2953.96 & Taboada & 24998.090 & 37863.210 & 39327.47\\
				Bergondo & 205.538 & 330.999 & 345.125 & Castroverde & 22586.410 & 34178.620 & 35492.41\\
				Cambre & 1318.166 & 2123.785 & 2214.635 & Friol & 16680.300 & 25160.910 & 26121.6\\
				Carral & 1254.754 & 2021.311 & 2107.982 & Guntín & 26225.670 & 39743.450 & 41280.47\\
				Culleredo & 495.855 & 798.269 & 832.388 & Lugo & 27777.350 & 42136.230 & 43771.97\\
				Oleiros & 103.212 & 166.133 & 173.227 & O Corgo & 9887.222 & 14864.750 & 15428.46\\
				Arzúa & 28410.440 & 46891.87 & 49002.49 & Outeiro de Rei & 9009.945 & 13538.360 & 14049.88\\
				Boimorto & 10174.060 & 16507.43 & 17224.87 & Portomarín & 15279.160 & 23032.430 & 23910.31\\
				O Pino & 6349.623 & 10269.96 & 10713.25 & Meira & 7054.810 & 10592.030 & 10991.64\\
				Touro & 13207.470 & 21486.01 & 22423.29 & Pol & 26242.340 & 39781.190 & 41322.81\\
				A Laracha & 7893.943 & 12782.57 & 13335.47 & Ribeira  de Piquín & 1090.563 & 1632.442 & 1693.474\\
				Cabana de Bergantiños & 5406.257 & 8736.76 & 9113.683 & Ríotorto & 5090.996 & 7637.931 & 7925.086\\
				Carballo & 6055.162 & 9792.728 & 10214.87 & As Nogais & 166.419 & 249.051 & 258.347\\
				Coristanco & 6232.419 & 10081.63 & 10516.04 & Baralla & 3375.138 & 5057.779 & 5247.408\\
				Laxe & 337.911 & 544.104 & 567.378 & Becerreá & 911.771 & 1365.346 & 1416.358\\
				Malpica de Bergantiños & 1379.753 & 2223.381 & 2318.361 & Pedrafita do Cebreiro & 1.522 & 2.278 & 2.363\\
				Ponteceso & 7685.086 & 12444.55 & 12983 & Láncara & 10238.260 & 15396.810 & 15977.87\\
				Aranga & 10551.900 & 17130.17 & 17873.98 & O Incio & 2828.965 & 4238.901 & 4397.895\\
				Betanzos & 113.820 & 183.238 & 191.053 & O Páramo & 12417.180 & 18695.760 & 19405.89\\
				Cesuras & 4995.667 & 8071.542 & 8418.723 & Paradela & 16951.010 & 25575.920 & 26551.46\\
				Coirós & 30.209 & 48.622 & 50.697 & Samos & 6645.102 & 9973.214 & 10348.85\\
				Curtis & 15328.460 & 24990.01 & 26086.61 & Sarria & 32652.450 & 49665.210 & 51610.25\\
				Irixoa & 5004.705 & 8083.079 & 8431.671 & Triacastela & 932.899 & 1396.759 & 1448.846\\
				Miño & 936.445 & 1508.965 & 1573.434 & A Pastoriza & 52836.040 & 81419.360 & 84710.7\\
				Oza dos Ríos & 1727.941 & 2784.591 & 2904.081 & Abadín & 7610.607 & 11430.920 & 11861.28\\
				Paderne & 1932.244 & 3113.961 & 3247.486 & Begonte & 3877.536 & 5812.512 & 6030.794\\
				Vilarmaior & 1293.701 & 2085.036 & 2174.23 & Castro de Rei & 43701.120 & 66932.050 & 69595.02\\
				Vilasantar & 6290.517 & 10175.45 & 10615.63 & Cospeito & 33246.670 & 50591.670 & 52574.8\\
				A Capela & 5145.646 & 8314.519 & 8672.351 & Guitiriz & 22991.550 & 34790.150 & 36132.18\\
				As Pontes & 1260.696 & 2031.264 & 2118.132 & Vilalba & 15007.370 & 22619.310 & 23479.44\\
				Cabanas & 444.224 & 715.214 & 745.777 & Xermade & 12086.220 & 18193.830 & 18883.89\\
				Monfero & 11961.690 & 19431.49 & 20278.44 & A Pobra de Brollón & 4929.655 & 7394.900 & 7672.226\\
				Pontedeume & 282.995 & 455.61 & 475.073 & Bóveda & 2641.195 & 3957.304 & 4105.905\\
				As Somozas & 552.911 & 890.663 & 928.743 & Monforte de Lemos & 3634.172 & 5446.519 & 5650.655\\
				Fene & 207.500 & 334.137 & 348.399 & O Saviñao & 14543.150 & 21909.520 & 22744.41\\
				Moeche & 1495.092 & 2409.048 & 2512.087 & Pantón & 734.865 & 1100.147 & 1141.217\\
				Mugardos & 188.553 & 303.576 & 316.577 & Sober & 618.255 & 925.283 & 959.884\\\cdashline{5-8}
				Narón & 620.359 & 998.848 & 1041.598 & \cellcolor{GGG}Lobios & 45.104 & 50.749 & 50.749\\
				Neda & 210.939 & 339.56 & 354.093 &\cellcolor{GGG} Baltar & 87.745 & 98.726 & 98.726\\
				Ferrol & 269.059 & 433.147 & 451.657 & \cellcolor{GGG}Rairiz de Veiga & 23.176 & 26.076 & 26.076\\
				San Sadurniño & 4409.964 & 7121.33 & 7427.985 & \cellcolor{GGG}Sandiás & 130.010 & 146.280 & 146.28\\
				Valdoviño & 540.499 & 870.278 & 907.536 &\cellcolor{GGG} Sarreaus & 1027.788 & 1156.415 & 1156.415\\
				Cee & 310.477 & 499.899 & 521.305 & \cellcolor{GGG}Trasmiras & 120.148 & 135.184 & 135.184\\
				Dumbría & 9230.228 & 14964.45 & 15613.74 & \cellcolor{GGG}Xinzo de Limia & 1898.470 & 2136.062 & 2136.062\\
				Muxía & 9465.518 & 15348.43 & 16015.48 & \cellcolor{GGG}Allariz & 1161.774 & 1307.170 & 1307.17\\
				Carnota & 22.014 & 35.437 & 36.955 & \cellcolor{GGG}Maceda & 2456.292 & 2763.695 & 2763.695\\
				Lousame & 2800.800 & 4516.836 & 4710.738 & \cellcolor{GGG}Xunqueira de Ambía & 576.508 & 648.658 & 648.658\\
				Noia & 54.490 & 87.685 & 91.421 & \cellcolor{GGG}Xunqueira de Espadanedo & 47.772 & 53.751 & 53.751\\
				Outes & 1023.983 & 1649.573 & 1720.158 &\cellcolor{GGG} O Carballiño & 43.059 & 48.448 & 48.448\\
				Rianxo & 12.160 & 19.571 & 20.409 & \cellcolor{GGG}O Irixo & 190.799 & 214.678 & 214.678\\
				Ribeira & 139.627 & 224.796 & 234.411 & \cellcolor{GGG}Piñor & 703.044 & 791.029 & 791.029\\
				Cerceda & 6558.788 & 10611.55 & 11069.3 & \cellcolor{GGG}San Amaro & 104.069 & 117.094 & 117.094\\
				Frades & 27393.980 & 45171.42 & 47195.98 & \cellcolor{GGG}San Cristovo de Cea & 368.480 & 414.595 & 414.595\\
				Mesía & 24147.090 & 39715.34 & 41485.98 & \cellcolor{GGG}Barbadás & 35.077 & 39.467 & 39.467\\
				Ordes & 15064.230 & 24552.17 & 25628.74 & \cellcolor{GGG}Coles & 138.718 & 156.078 & 156.078\\
				Oroso & 4858.644 & 7849.281 & 8186.894 & \cellcolor{GGG}Taboadela & 120.883 & 136.011 & 136.011\\
				Tordoia & 21102.820 & 34604.67 & 36142.84 & \cellcolor{GGG}A Teixeira & 216.548 & 243.649 & 243.649\\
				Trazo & 20326.700 & 33323.01 & 34794.6 & \cellcolor{GGG}A Bola & 8.833 & 9.938 & 9.938\\
				Cerdido & 664.400 & 1069.825 & 1115.532 & \cellcolor{GGG}A Merca & 21.056 & 23.692 & 23.692\\
				Mañón & 170.024 & 273.731 & 285.43 & \cellcolor{GGG}Cartelle & 309.888 & 348.670 & 348.67\\
				Ortigueira & 1786.389 & 2879.073 & 3002.295 & \cellcolor{GGG}Celanova & 294.651 & 331.527 & 331.527\\
				Ames & 1115.917 & 1797.564 & 1874.654 & \cellcolor{GGG}Ramirás & 34.622 & 38.955 & 38.955\\
				Boqueixón & 7677.820 & 12428.27 & 12967.14 & \cellcolor{GGG}Manzaneda & 79.215 & 89.128 & 89.128\\
				Brión & 7363.648 & 11920.24 & 12434.66 & \cellcolor{GGG}Río & 192.341 & 216.413 & 216.413\\
				Santiago de Compostela & 7114.561 & 11515.45 & 12011.24 & \cellcolor{GGG}Riós & 746.349 & 839.754 & 839.754\\
				Teo & 1121.647 & 1807.371 & 1884.694 & \cellcolor{GGG}A Gudiña & 176.590 & 198.690 & 198.69\\
				Val do Dubra & 12802.350 & 20822.55 & 21729.74 &\cellcolor{GGG} A Mezquita & 536.833 & 604.018 & 604.018\\
				Vedra & 1959.834 & 3158.399 & 3293.903 & \cellcolor{GGG}Viana do Bolo & 891.628 & 1003.214 & 1003.214\\
				Padrón & 788.590 & 1270.042 & 1324.383 & \cellcolor{GGG}Vilariño de Conso & 100.865 & 113.489 & 113.489\\\cdashline{5-8}
				Rois & 7579.308 & 12271.62 & 12803.2 & \cellcolor{GGG}Caldas de Reis & 3.958 & 4.406 & 5.083\\
				Melide & 8838.328 & 14325.72 & 14946.73 & \cellcolor{GGG}Cuntis & 558.563 & 621.781 & 717.134\\
				Santiso & 9942.311 & 16127.35 & 16829.41 &\cellcolor{GGG} Moraña & 27.118 & 30.190 & 34.821\\
				Sobrado & 6740.872 & 10904.44 & 11374.99 &\cellcolor{GGG} Portas & 59.908 & 66.701 & 76.935\\
				Toques & 3652.379 & 5895.82 & 6148.569 & \cellcolor{GGG}Valga & 227.073 & 252.767 & 291.566\\
				Vimianzo & 12901.350 & 20974.28 & 21891.31 &\cellcolor{GGG} Agolada & 6696.345 & 7428.076 & 8589.148\\
				Zas & 9872.865 & 16016.5 & 16710.27 & \cellcolor{GGG}Dozón & 10485.720 & 11634.180 & 13471.12\\
				Mazaricos & 34047.980 & 56401.44 & 58975.79 & \cellcolor{GGG}Lalín & 54898.530 & 57527.770 & 71488.28\\
				Santa Comba & 37020.740 & 61457.6 & 64269.84 & \cellcolor{GGG}Rodeiro & 32107.390 & 34666.550 & 41533.86\\\cdashline{1-4}
				A Fonsagrada & 874.827 & 1309.469 & 1358.573 & \cellcolor{GGG}Silleda & 34733.990 & 37337.680 & 44897.17\\
				Baleira & 6356.899 & 9540.247 & 9900.572 & \cellcolor{GGG}Vila de Cruces & 10356.950 & 11494.110 & 13307.5\\
				Alfoz & 753.410 & 1127.976 & 1170.209 & \cellcolor{GGG}Meis & 19.677 & 21.912 & 25.271\\
				Foz & 2311.133 & 3462.34 & 3591.597 & \cellcolor{GGG}O Grove & 17.860 & 19.881 & 22.938\\
				Lourenzá & 2723.323 & 4080.861 & 4233.913 & \cellcolor{GGG}Pontecesures & 5.383 & 5.994 & 6.913\\
				Mondoñedo & 2878.948 & 4312.787 & 4474.662 & \cellcolor{GGG}Ribadumia & 88.041 & 98.006 & 113.065\\
				O Valadouro & 1624.842 & 2433.864 & 2524.828 & \cellcolor{GGG}Campo Lameiro & 33.732 & 37.559 & 43.324\\
				Ourol & 14.010 & 20.966 & 21.748 & \cellcolor{GGG}Cotobade & 146.652 & 163.235 & 188.305\\
				A Pontenova & 3165.579 & 4742.729 & 4920.772 &\cellcolor{GGG} Poio & 12.948 & 14.417 & 16.627\\
				Barreiros & 10940.310 & 16452.56 & 17076.06 & \cellcolor{GGG}Pontevedra & 204.923 & 228.146 & 263.133\\
				Ribadeo & 16964.190 & 25593.94 & 26570.72 & \cellcolor{GGG}A Estrada & 12979.930 & 14402.690 & 16681.29\\
				Trabada & 10638.900 & 16001.62 & 16607.3 & \cellcolor{GGG}Cerdedo & 44.601 & 49.652 & 57.276\\
				Antas de Ulla & 2931.569 & 4392.703 & 4557.409 & \cellcolor{GGG}Forcarei & 8311.015 & 9222.576 & 10668.92\\\hline					
		\end{tabular}}
		\caption{Estimated $RA-$random arrival rule for $190$ councils in Galicia with provinces.}\label{RA_190concellos_provincias}
	\end{center}
\end{table}

\newpage
 \begin{table}[h!]
 	\begin{center}
		\resizebox{0.69\textwidth}{!}{	
 			\begin{tabular}{|p{0.28\textwidth}|p{0.10\textwidth}|p{0.10\textwidth}|p{0.10\textwidth}||p{0.28\textwidth}|p{0.10\textwidth}|p{0.10\textwidth}|p{0.10\textwidth}|}\hline
 				Council & $\rho=40$ & $\rho=10$ & $\rho=5$ & Council & $\rho=40$ & $\rho=10$ & $\rho=5$ \\\hline
 		A Baña & 11451.110 & 16955.530 & 17822.120 & Monterroso & 10944.430 & 16201.870 & 17030.740\\
 		Negreira & 15791.030 & 23387.560 & 24573.250 & Palas de Rei & 15799.280 & 23392.120 & 24578.100\\\cdashline{1-4}\cdashline{5-8}
 		A  Coruña & 360.288 & 531.140 & 558.457 & Carballedo & 14209.060 & 21171.920 & 22450.000\\
 		Abegondo & 1756.319 & 2589.130 & 2722.369 & Chantada & 32556.160 & 48500.960 & 51324.440\\
 		Arteixo & 1831.884 & 2700.394 & 2839.429 & Taboada & 24353.730 & 36285.790 & 38406.980\\\cdashline{5-8}
 		Bergondo & 214.102 & 315.637 & 331.882 & \cellcolor{GGG}Castroverde & 22160.530 & 33429.880 & 35584.400\\
 		Cambre & 1373.688 & 2025.089 & 2129.306 &\cellcolor{GGG} Friol & 16328.200 & 24636.370 & 26203.130\\
 		Carral & 1307.439 & 1927.410 & 2026.596 & \cellcolor{GGG}Guntín & 25755.650 & 38850.940 & 41364.770\\
 		Culleredo & 516.358 & 761.232 & 800.401 & \cellcolor{GGG}Lugo & 27299.330 & 41185.480 & 43859.130\\
 		Oleiros & 107.474 & 158.443 & 166.594 & \cellcolor{GGG}O Corgo & 9658.551 & 14572.950 & 15493.000\\\cdashline{1-4}
 		Arzúa & 30355.080 & 45146.060 & 47630.390 &\cellcolor{GGG} Outeiro de Rei & 8797.599 & 13273.300 & 14110.220\\
 		Boimorto & 10721.570 & 15949.860 & 16825.610 &\cellcolor{GGG} Portomarín & 14953.020 & 22558.600 & 23992.910\\\cdashline{5-8}
 		O Pino & 6676.904 & 9931.684 & 10479.330 & Meira & 6808.943 & 10086.130 & 10617.180\\
 		Touro & 13948.35 & 20747.52 & 21889.040 & Pol & 25489.050 & 37776.160 & 39721.350\\\cdashline{1-4}
 		A Laracha & 8285.568 & 12279.320 & 12912.890 & Ribeira de Piquín & 1050.212 & 1555.509 & 1638.070\\
 		Cabana de Bergantiños & 5666.316 & 8396.605 & 8830.473 & Ríotorto & 4910.564 & 7273.834 & 7656.334\\\cdashline{5-8}
 		Carballo & 6349.184 & 9409.005 & 9895.639 & As Nogais & 159.678 & 235.204 & 247.256\\
 		Coristanco & 6536.639 & 9686.969 & 10187.340 & Baralla & 3241.979 & 4775.431 & 5019.883\\
 		Laxe & 353.146 & 523.259 & 550.508 & Becerreá & 875.300 & 1289.280 & 1355.274\\
 		Malpica de Bergantiños & 1442.675 & 2137.650 & 2248.645 & Pedrafita do Cebreiro & 1.461 & 2.152 & 2.262\\\cdashline{5-8}
 		Ponteceso & 8066.445 & 11954.330 & 12571.750 & Láncara & 9938.288 & 14831.620 & 15792.000\\\cdashline{1-4}
 		Aranga & 11108.890 & 16499.180 & 17386.810 & O Incio & 2739.691 & 4088.900 & 4353.577\\
 		Betanzos & 119.045 & 176.751 & 186.335 & O Páramo & 12066.200 & 18007.590 & 19172.840\\
 		Cesuras & 5238.767 & 7779.560 & 8201.287 & Paradela & 16491.090 & 24614.240 & 26176.950\\
 		Coirós & 31.590 & 46.908 & 49.453 & Samos & 6442.074 & 9613.609 & 10232.000\\
 		Curtis & 16193.470 & 24053.850 & 25338.880 & Sarria & 31947.240 & 47655.230 & 50567.910\\
 		Irixoa & 5247.926 & 7793.034 & 8215.021 &\cellcolor{GGG} Triacastela & 902.742& 1347.355 & 1434.815\\\cdashline{5-8}
 		Miño & 980.170 & 1455.504 & 1534.401 & \cellcolor{GGG}A Pastoriza & 53029.390 & 81768.330 & 85437.010\\
 		Oza dos Ríos & 1808.735 & 2686.055 & 2831.741 &\cellcolor{GGG} Abadín & 7505.611 & 11548.110 & 12023.040\\
 		Paderne & 2022.789 & 3003.704 & 3166.417 & \cellcolor{GGG}Begonte & 3818.848 & 5875.223 & 6115.180\\
 		Vilarmaior & 1354.222 & 2011.009 & 2120.041 & \cellcolor{GGG}Castro de Rei & 43673.860 & 67332.990 & 70261.060\\
 		Vilasantar & 6604.351 & 9807.491 & 10338.170 & \cellcolor{GGG}Cospeito & 33081.210 & 50967.880 & 53146.630\\\cdashline{1-4}
 		A Capela & 5380.326 & 7952.533 & 8357.406 & \cellcolor{GGG}Guitiriz & 22786.760 & 35084.820 & 36563.670\\
 		As Pontes & 1315.442 & 1944.272 & 2043.247 &\cellcolor{GGG} Vilalba & 14832.880 & 22830.990 & 23775.300\\
 		Cabanas & 463.164 & 684.594& 719.460 & \cellcolor{GGG}Xermade & 11938.330 & 18369.680 & 19127.110\\\cdashline{5-8}
 		Monfero & 12559.880 & 18572.700 & 19509.070 & A Pobra de Brollón & 4746.417 & 7023.104 & 7381.346\\
 		Pontedeume & 295.063& 436.146 & 458.349 & Bóveda & 2540.896 & 3760.058 & 3951.856\\\cdashline{1-4}
 		As Somozas & 575.818 & 849.498 & 893.210 & Monforte de Lemos & 3496.737 & 5173.676 & 5437.639\\
 		Fene & 216.041 & 318.737 & 335.127 & O Saviñao & 14043.650 & 20789.900 & 21836.560\\
 		Moeche & 1557.234 & 2297.469 & 2415.619 & Pantón & 706.564& 1045.393 & 1098.932\\
 		Mugardos & 196.307 & 289.608 & 304.505 & Sober & 594.382 & 879.458 & 924.490\\\cdashline{5-8}
 		Narón & 645.807 & 952.781 & 1001.791 & Lobios & 40.166 & 59.134 & 62.177\\\cdashline{5-8}
 		Neda & 219.574 & 323.944 & 340.610 & Baltar & 78.127 & 115.103 & 121.016\\
 		Ferrol & 280.069 & 413.212 & 434.453 & Rairiz de Veiga & 20.636& 30.401 & 31.964\\
 		San Sadurniño & 4601.851 & 6788.182 & 7137.990 & Sandiás & 115.759 & 170.546 & 179.308\\
 		Valdoviño & 562.630 & 830.081 & 872.786 & Sarreaus & 915.125 & 1348.274 & 1417.525\\\cdashline{1-4}
 		Cee & 323.687 & 478.632 & 502.895 & Trasmiras & 106.978& 157.607 & 165.707\\
 		Dumbría & 9675.472 & 14308.650 & 15030.970 & Xinzo de Limia & 1690.394 & 2490.401 & 2618.415\\\cdashline{5-8}
 		Muxía & 9924.243 & 14676.620 & 15418.180 & Allariz & 1035.027 & 1524.455 & 1602.605\\\cdashline{1-4}
 		Carnota & 22.897 & 33.723 & 35.462 & Maceda & 2188.291 & 3223.063 & 3388.339\\\cdashline{1-4}
 		Lousame & 2918.827 & 4300.896 & 4521.826 & Xunqueira de Ambía & 513.615 & 756.471 & 795.267\\
 		Noia & 56.687 & 83.525 & 87.817 & Xunqueira de Espadanedo & 42.560 & 62.685 & 65.899\\\cdashline{5-8}
 		Outes & 1066.153 & 1570.945 & 1651.591 & O Carballiño & 38.336 & 56.455 & 59.359\\\cdashline{1-4}
 		Rianxo & 12.648 & 18.624 & 19.583 & O Irixo & 169.873 & 250.158 & 263.026\\
 		Ribeira & 145.276 & 213.920 & 224.931 & Piñor & 625.935 & 921.765 & 969.172\\\cdashline{1-4}
 		\cellcolor{GGG}Cerceda & 6959.964 & 10511.820 & 11166.470 & San Amaro & 92.656 & 136.444 & 143.466\\
 		\cellcolor{GGG}Frades & 29500.290 & 44555.370 & 47436.160 & San Cristovo de Cea & 328.066 & 483.111 & 507.968\\\cdashline{5-8}
 		\cellcolor{GGG}Mesía & 25960.570 & 39202.110 & 41716.570 & Barbadás & 31.223 & 45.973 & 48.343\\
 		\cellcolor{GGG}Ordes & 16075.620 & 24282.480 & 25805.340 & Coles & 123.477 & 181.809 & 191.180\\
 		\cellcolor{GGG}Oroso & 5149.387 & 7777.363 & 8260.124 & Taboadela & 107.602 & 158.433& 166.599\\\cdashline{5-8}
 		\cellcolor{GGG}Tordoia & 22637.310 & 34192.250 & 36368.690 & A Teixeira & 192.825 & 283.864 & 298.511\\\cdashline{5-8}
 		\cellcolor{GGG}Trazo & 21794.790 & 32917.390 & 35014.720 & A Bola & 7.866 & 11.580 & 12.177\\\cdashline{1-4}
 		Cerdido & 691.389 & 1018.574 & 1070.979 & A Merca & 18.751 & 27.606 & 29.028\\
 		Mañón & 176.927 & 260.656 & 274.067 & Cartelle & 275.954 & 406.273 & 427.207\\
 		Ortigueira & 1860.196 & 2740.382 & 2881.516 & Celanova & 262.385 & 386.299 & 406.201\\\cdashline{1-4}
 		Ames & 1166.492 & 1730.091 & 1821.595 & Ramirás & 30.831 & 45.391 & 47.729\\\cdashline{5-8}
 		Boqueixón & 8056.317 & 11949.730 & 12576.550 & Manzaneda & 70.551 & 103.819 & 109.162\\
 		Brión & 7726.275 & 11460.320 & 12061.270 & Río & 171.304 & 252.085 & 265.058\\\cdashline{5-8}
 		Santiago de Compostela & 7463.847 & 11070.290 & 11650.810 & Riós & 664.556 & 978.553 & 1028.845\\\cdashline{5-8}
 		Teo & 1172.661 & 1739.139 & 1831.231 & A Gudiña & 157.243 & 231.561 & 243.472\\
 		Val do Dubra & 13482.120 & 19999.510 & 21042.080 & A Mezquita & 478.019 & 703.939 & 740.151\\
 		Vedra & 2049.166 & 3038.996 & 3199.641 & Viana do Bolo & 793.949 & 1169.183 & 1229.303\\\cdashline{1-4}
 		Padrón & 821.185 & 1211.911 & 1273.980 & Vilariño de Conso & 89.814 & 132.264 & 139.067\\\cdashline{5-8}
 		Rois & 7924.836 & 11694.050 & 12294.850 & Caldas de Reis & 3.524 & 5.188& 5.455\\\cdashline{1-4}
 		Melide & 9272.134 & 13729.680 & 14431.820 & Cuntis & 497.262 & 732.144 & 769.776\\
 		Santiso & 10438.320 & 15456.210 & 16246.410 & Moraña & 24.142 & 35.545 & 37.372\\
 		Sobrado & 7061.695 & 10454.940 & 10992.470 & Portas & 53.334 & 78.526 & 82.562\\
 		Toques & 3819.443 & 5655.087 & 5946.622 & Valga & 202.153 & 297.640 & 312.940\\\cdashline{1-4}\cdashline{5-8}
 		Vimianzo & 13559.310 & 20062.440 & 21075.880 &\cellcolor{GGG} Agolada & 6061.018 & 9181.164 & 9714.568\\
 		Zas & 10356.960 & 15320.970 & 16100.120 & \cellcolor{GGG}Dozón & 9491.347 & 14372.140 & 15218.970\\\cdashline{1-4}
 		Mazaricos & 36548.380 & 54469.950 & 57821.430 & \cellcolor{GGG}Lalín & 49688.000 & 75187.950 & 80046.840\\
 		Santa Comba & 39792.030 & 59304.370 & 62934.520 & \cellcolor{GGG}Rodeiro & 29059.270 & 43994.400 & 46680.930\\\cdashline{1-4}
 		A Fonsagrada & 839.879 & 1237.927 & 1301.450 & \cellcolor{GGG}Silleda & 31436.990 & 47589.920 & 50510.720\\
 		Baleira & 6114.512 & 9008.755 & 9472.705 & \cellcolor{GGG}Vila de Cruces & 9374.416 & 14197.470 & 15032.310\\\cdashline{1-4}\cdashline{5-8}
 		Alfoz & 723.360 & 1067.157 & 1121.690 & Meis & 17.520 & 25.788 & 27.116\\
 		Foz & 2219.471 & 3274.134 & 3441.811 & O Grove & 15.903 & 23.407 & 24.613\\
 		Lourenzá & 2615.857 & 3858.706 & 4056.351 & Pontecesures & 4.794 & 7.055 & 7.419\\
 		Mondoñedo & 2765.104 & 4078.798 & 4287.650 & Ribadumia & 78.394 & 115.385 & 121.330\\\cdashline{5-8}
 		O Valadouro & 1560.199 & 2301.602 & 2419.380 & Campo Lameiro & 30.018 & 44.211 & 46.489\\\cdashline{1-4}
 		Ourol & 13.434 & 19.777 & 20.797 & Cotobade & 130.503& 192.212 & 202.115\\\cdashline{1-4}
 		A Pontenova & 3049.520 & 4521.060 & 4760.084 & Poio & 11.522 & 16.970& 17.845\\
 		Barreiros & 10565.600 & 15667.080 & 16488.780 & Pontevedra & 182.358 & 268.584 & 282.425\\\cdashline{5-8}
 		Ribadeo & 16417.600 & 24350.330 & 25616.730 & A Estrada & 11579.860 & 17112.460 & 17977.110\\
 		Trabada & 10276.060 & 15238.350 & 16037.640 & Cerdedo & 39.790 & 58.783 & 61.776\\\cdashline{1-4}
 		Antas de Ulla & 2821.933 & 4176.477 & 4392.507 & Forcarei & 7414.424 & 10954.070 & 11510.510\\
 				\hline
 		\end{tabular}}
 		\caption{Estimated $RA-$random arrival rule for $190$ councils in Galicia with counties.}\label{RA_190concellos_counties}
 	\end{center}
 \end{table}

\begin{table}[h!]
	\begin{center}
	\resizebox{0.725\textwidth}{!}{	
	\begin{tabular}{|p{0.28\textwidth}|p{0.18\textwidth}|p{0.18\textwidth}||p{0.28\textwidth}|p{0.18\textwidth}|p{0.18\textwidth}|}\hline
		Council      &       Talmud rule     &   Proport. rule     &   Council     &      Talmud rule     &   Proport. rule       \\\hline
A Baña&9626.5735&11551.8882&Monterroso&9193.6270&11032.3524\\
Negreira&13542.3638&15931.0464&Palas de Rei&13417.4335&15924.8442\\
A Coruña&303.3365&364.0038&Carballedo&11878.9765&14254.7718\\
Abegondo&1478.7530&1774.5036&Chantada&41314.7705&32663.2464\\
Arteixo&1542.3180&1850.7816&Taboada&27596.8595&24432.4998\\
Bergondo&180.2660&216.3192&Castroverde&23645.8575&22061.8986\\
Cambre&1156.5770&1387.8924&Friol&13973.0155&16258.1934\\
Carral&1100.7945&1320.9534&Guntín&29611.5545&25641.3168\\
Culleredo&434.7525&521.7030&Lugo&32175.0025&27179.3856\\
Oleiros&90.4880&108.5856&O Corgo&8013.5865&9616.3038\\
Arzúa&37801.6708&30486.6306&Outeiro de Rei&7299.261&8759.1132\\
Boimorto&8974.6185&10769.5422&Portomarín&12405.9045&14887.0854\\
O Pino&5588.2305&6705.8766&Meira&5712.7630&6855.3156\\
Touro&11674.4645&14009.3574&Pol&29650.1155&25664.4534\\
A Laracha&6952.5755&8343.0906&Ribeira de Piquín&881.1390&1057.3668\\
Cabana de Bergantiños&4754.6185&5705.5422&Ríotorto&4119.9000&4943.8800\\
Carballo&5327.9700&6393.5640&As Nogais&134.4355&161.3226\\
Coristanco&5485.2220&6582.2664&Baralla&2729.3980&3275.2776\\
Laxe&296.3445&355.6134&Becerreá&736.8945&884.2734\\
Malpica de Bergantiños&1210.5585&1452.6702&Pedrafita do Cebreiro&1.2300&1.476\\
Ponteceso&6768.9880&8122.7856&Láncara&8298.0210&9957.6252\\
Aranga&9313.4890&11176.1868&O Incio&2287.5670&2745.0804\\
Betanzos&99.7915&119.7498&O Páramo&10073.9650&12088.7580\\
Cesuras&4392.1640&5270.5968&Paradela&14417.0995&16524.6438\\
Coirós&26.4830&31.7796&Samos&5378.9125&6454.6950\\
Curtis&14143.1628&16291.5258&Sarria&40221.5915&32007.3390\\
Irixoa&4399.5770&5279.4924&Triacastela&753.8320&904.5984\\
Miño&821.7140&986.0568&A Pastoriza&73979.4395&52262.0478\\
Oza dos Ríos&1516.4425&1819.7310&Abadín&6164.2550&7397.1060\\
Paderne&1695.7235&2034.8682&Begonte&3136.5180&3763.8216\\
Vilarmaior&1135.3645&1362.4374&Castro de Rei&58612.1605&43041.6804\\
Vilasantar&5536.8375&6644.2050&Cospeito&41212.0505&32601.6144\\
A Capela&4524.7715&5429.7258&Guitiriz&24305.1695&22457.4858\\
As Pontes&1106.2095&1327.4514&Vilalba&12182.1930&14618.6316\\
Cabanas&389.5165&467.4198&Xermade&9803.7680&11764.5216\\
Monfero&10562.5925&12675.1110&A Pobra de Brollón&3989.1025&4786.9230\\
Pontedeume&248.1515&297.7818&Bóveda&2135.6215&2562.7458\\
As Somozas&485.0310&582.0372&Monforte de Lemos&2938.7360&3526.4832\\
Fene&181.9810&218.3772&O Saviñao&11802.7535&14163.3042\\
Moeche&1311.7040&1574.0448&Pantón&593.8240&712.5888\\
Mugardos&165.3555&198.4266&Sober&499.5595&599.4714\\\cdashline{4-6}
Narón&543.9955&652.7946&Lobios&33.8370&40.6044\\
Neda&184.9575&221.9490&Baltar&65.8265&78.9918\\
Ferrol&235.9155&283.0986&Rairiz de Veiga&17.3865&20.8638\\
San Sadurniño&3876.2740&4651.5288&Sandiás&97.5335&117.0402\\
Valdoviño&473.9345&568.7214&Sarreaus&771.0490&925.2588\\
Cee&272.2555&326.7066&Trasmiras&90.1350&108.1620\\
Dumbría&8137.8825&9765.4590&Xinzo de Limia&1424.2360&1709.0832\\
Muxía&8347.2725&10016.7270&Allariz&871.5655&1045.8786\\
Carnota&19.3015&23.1618&Maceda&1842.7150&2211.2580\\
Lousame&2459.2115&2951.0538&Xunqueira de Ambía&432.4975&518.9970\\
Noia&47.7605&57.3126&Xunqueira de Espadanedo&35.8390&43.0068\\
Outes&898.2555&1077.9066&O Carballiño&32.3030&38.7636\\
Rianxo&10.6600&12.7920&O Irixo&143.1380&171.7656\\
Ribeira&122.4445&146.9334&Piñor&527.4250&632.9100\\
Cerceda&5773.5680&6928.2816&San Amaro&78.0730&93.6876\\
Frades&35940.7768&29370.0942&San Cristovo de Cea&276.4345&331.7214\\
Mesía&30057.1438&25839.9144&Barbadás&26.3150&31.5780\\
Ordes&13664.8328&16004.5278&Coles&104.0665&124.8798\\
Oroso&4271.7105&5126.0526&Taboadela&90.6865&108.8238\\
Tordoia&24550.6688&22536.0294&A Teixeira&162.4550&194.9460\\
Trazo&23152.5268&21697.1442&A Bola&6.6265&7.9518\\
Cerdido&582.6325&699.1590&A Merca&15.7965&18.9558\\
Mañón&149.0985&178.9182&Cartelle&232.4785&278.9742\\
Ortigueira&1567.5875&1881.1050&Celanova&221.0480&265.2576\\
Ames&979.061&1174.8732&Ramirás&25.9735&31.1682\\
Boqueixón&6761.6925&8114.0310&Manzaneda&59.4270&71.3124\\
Brión&6484.6230&7781.5476&Río&144.2950&173.1540\\
Santiago de Compostela&6264.0365&7516.8438&Riós&559.9125&671.8950\\
Teo&984.1815&1181.0178&A Gudiña&132.4780&158.9736\\
Val do Dubra&11315.1825&13578.2190&A Mezquita&402.7335&483.2802\\
Vedra&1719.7775&2063.7330&Viana do Bolo&668.9010&802.6812\\
Padrón&691.6885&830.0262&Vilariño de Conso&75.6695&90.8034\\\cdashline{4-6}
Rois&6675.1840&8010.2208&Caldas de Reis&2.9690&3.5628\\
Melide&7789.9880&9347.9856&Cuntis&418.9665&502.7598\\
Santiso&8769.6145&10523.5374&Moraña&20.3405&24.4086\\
Sobrado&5932.7725&7119.3270&Portas&44.936&53.9232\\
Toques&3208.9905&3850.7886&Valga&170.3230&204.3876\\
Vimianzo&11399.9495&13679.9394&Agolada&5022.7920&6027.3504\\
Zas&8707.7510&10449.3012&Dozón&7865.1220&9438.1464\\
Mazaricos&48036.6838&36627.6384&Lalín&41178.2720&49413.9264\\
Santa Comba&53455.8028&39879.1098&Rodeiro&24083.1025&28899.7230\\\cdashline{1-3}
A Fonsagrada&706.8835&848.2602&Silleda&26053.255&31263.9060\\
Baleira&5146.0950&6175.3140&Vila de Cruces&7768.5380&9322.2456\\
Alfoz&608.9420&730.7304&Meis&14.7590&17.7108\\
Foz&1868.4500&2242.1400&O Grove&13.3965&16.0758\\
Lourenzá&2202.0510&2642.4612&Pontecesures&4.0380&4.8456\\
Mondoñedo&2327.6815&2793.2178&Ribadumia&66.0380&79.2456\\
O Valadouro&1313.4305&1576.1166&Campo Lameiro&25.3015&30.3618\\
Ourol&11.3190&13.5828&Cotobade&110.0005&132.0006\\
A Pontenova&2559.4450&3071.3340&Poio&9.7120&11.6544\\
Barreiros&8867.9365&10641.5238&Pontevedra&153.7085&184.4502\\
Ribadeo&14437.9095&16537.1298&A Estrada&9735.9840&11683.1808\\
Trabada&8625.3815&10350.4578&Cerdedo&33.4540&40.1448\\
Antas de Ulla&2370.3995&2844.4794&Forcarei&6233.9230&7480.7076\\
\hline
		\end{tabular}}
		\caption{New systems of milk quotas for $190$ councils in Galicia for the case $\rho=40\%$ with the provinces.}\label{rules_190concellos}
	\end{center}
\end{table}

\begin{table}[h!]
	\begin{center}
	\resizebox{0.725\textwidth}{!}{	
	\begin{tabular}{|p{0.28\textwidth}|p{0.18\textwidth}|p{0.18\textwidth}||p{0.28\textwidth}|p{0.18\textwidth}|p{0.18\textwidth}|}\hline
				Council      &       Talmud rule     &   Proport. rule     &   Council     &      Talmud rule     &   Proport. rule       \\\hline
A Baña&9626.5735&11551.8882&Monterroso&9193.6270&11032.3524\\
Negreira&13275.8720&15931.0464&Palas de Rei&13270.7035&15924.8442\\\cdashline{1-3}\cdashline{4-6}
A Coruña&303.3365&364.0038&Carballedo&11878.9765&14254.7718\\
Abegondo&1478.7530&1774.5036&Chantada&27219.3720&32663.2464\\
Arteixo&1542.3180&1850.7816&Taboada&20360.4165&24432.4998\\\cdashline{4-6}
Bergondo&180.2660&216.3192&Castroverde&26922.9694&22061.8986\\
Cambre&1156.5770&1387.8924&Friol&17250.1274&16258.1934\\
Carral&1100.7945&1320.9534&Guntín&32888.6664&25641.3168\\
Culleredo&434.7525&521.7030&Lugo&35452.1144&27179.3856\\
Oleiros&90.4880&108.5856&O Corgo&8013.5865&9616.3038\\\cdashline{1-3}
Arzúa&25405.5255&30486.6306&Outeiro de Rei&7299.2610&8759.1132\\
Boimorto&8974.6185&10769.5422&Portomarín&14964.9474&14887.0854\\\cdashline{4-6}
O Pino&5588.2305&6705.8766&Meira&5712.7630&6855.3156\\
Touro&11674.4645&14009.3574&Pol&21387.0445&25664.4534\\\cdashline{1-3}
A Laracha&6952.5755&8343.0906&Ribeira de Piquín&881.1390&1057.3668\\
Cabana de Bergantiños&4754.6185&5705.5422&Ríotorto&4119.9000&4943.8800\\\cdashline{4-6}
Carballo&5327.9700&6393.5640&As Nogais&134.4355&161.3226\\
Coristanco&5485.2220&6582.2664&Baralla&2729.3980&3275.2776\\
Laxe&296.3445&355.6134&Becerreá&736.8945&884.2734\\
Malpica de Bergantiños&1210.5585&1452.6702&Pedrafita do Cebreiro&1.2300&1.4760\\\cdashline{4-6}
Ponteceso&6768.9880&8122.7856&Láncara&8298.0210&9957.6252\\\cdashline{1-3}
Aranga&9313.4890&11176.1868&O Incio&2287.5670&2745.0804\\
Betanzos&99.7915&119.7498&O Páramo&10073.9650&12088.7580\\
Cesuras&4392.1640&5270.5968&Paradela&13770.5365&16524.6438\\
Coirós&26.4830&31.7796&Samos&5378.9125&6454.6950\\
Curtis&13576.2715&16291.5258&Sarria&29361.2433&32007.3390\\
Irixoa&4399.5770&5279.4924&Triacastela&753.8320&904.5984\\\cdashline{4-6}
Miño&821.7140&986.0568&A Pastoriza&77895.6826&52262.0478\\
Oza dos Ríos&1516.4425&1819.7310&Abadín&6164.2550&7397.1060\\
Paderne&1695.7235&2034.8682&Begonte&3136.5180&3763.8216\\
Vilarmaior&1135.3645&1362.4374&Castro de Rei&62528.4036&43041.6804\\
Vilasantar&5536.8375&6644.2050&Cospeito&45128.2936&32601.6144\\\cdashline{1-3}
A Capela&4524.7715&5429.7258&Guitiriz&28221.4126&22457.4858\\
As Pontes&1106.2095&1327.4514&Vilalba&15156.6556&14618.6316\\
Cabanas&389.5165&467.4198&Xermade&10399.8056&11764.5216\\\cdashline{4-6}
Monfero&10562.5925&12675.1110&A Pobra de Brollón&3989.1025&4786.9230\\
Pontedeume&248.1515&297.7818&Bóveda&2135.6215&2562.7458\\\cdashline{1-3}
As Somozas&485.0310&582.0372&Monforte de Lemos&2938.7360&3526.4832\\
Fene&181.9810&218.3772&O Saviñao&11802.7535&14163.3042\\
Moeche&1311.7040&1574.0448&Pantón&593.8240&712.5888\\
Mugardos&165.3555&198.4266&Sober&499.5595&599.4714\\\cdashline{4-6}
Narón&543.9955&652.7946&Lobios&33.8370&40.6044\\\cdashline{4-6}
Neda&184.9575&221.9490&Baltar&65.8265&78.9918\\
Ferrol&235.9155&283.0986&Rairiz de Veiga&17.3865&20.8638\\
San Sadurniño&3876.2740&4651.5288&Sandiás&97.5335&117.0402\\
Valdoviño&473.9345&568.7214&Sarreaus&771.0490&925.2588\\\cdashline{1-3}
Cee&272.2555&326.7066&Trasmiras&90.1350&108.1620\\
Dumbría&8137.8825&9765.4590&Xinzo de Limia&1424.2360&1709.0832\\\cdashline{4-6}
Muxía&8347.2725&10016.7270&Allariz&871.5655&1045.8786\\\cdashline{1-3}
Carnota&19.3015&23.1618&Maceda&1842.7150&2211.2580\\\cdashline{1-3}
Lousame&2459.2115&2951.0538&Xunqueira de Ambía&432.4975&518.9970\\
Noia&47.7605&57.3126&Xunqueira de Espadanedo&35.8390&43.0068\\\cdashline{4-6}
Outes&898.2555&1077.9066&O Carballiño&32.3030&38.7636\\\cdashline{1-3}
Rianxo&10.6600&12.7920&O Irixo&143.1380&171.7656\\
Ribeira&122.4445&146.9334&Piñor&527.4250&632.9100\\\cdashline{1-3}
Cerceda&5773.5680&6928.2816&San Amaro&78.0730&93.6876\\
Frades&38049.7816&29370.0942&San Cristovo de Cea&276.4345&331.7214\\\cdashline{4-6}
Mesía&32166.1486&25839.9144&Barbadás&26.3150&31.5780\\
Ordes&15773.8376&16004.5278&Coles&104.0665&124.8798\\
Oroso&4271.7105&5126.0526&Taboadela&90.6865&108.8238\\\cdashline{4-6}
Tordoia&26659.6736&22536.0294&A Teixeira&162.4550&194.9460\\\cdashline{4-6}
Trazo&25261.5316&21697.1442&A Bola&6.6265&7.9518\\\cdashline{1-3}
Cerdido&582.6325&699.1590&A Merca&15.7965&18.9558\\
Mañón&149.0985&178.9182&Cartelle&232.4785&278.9742\\
Ortigueira&1567.5875&1881.1050&Celanova&221.0480&265.2576\\\cdashline{1-3}
Ames&979.0610&1174.8732&Ramirás&25.9735&31.1682\\\cdashline{4-6}
Boqueixón&6761.6925&8114.0310&Manzaneda&59.4270&71.3124\\
Brión&6484.6230&7781.5476&Río&144.2950&173.1540\\\cdashline{4-6}
Santiago de Compostela&6264.0365&7516.8438&Riós&559.9125&671.8950\\\cdashline{4-6}
Teo&984.1815&1181.0178&A Gudiña&132.4780&158.9736\\
Val do Dubra&11315.1825&13578.2190&A Mezquita&402.7335&483.2802\\
Vedra&1719.7775&2063.7330&Viana do Bolo&668.9010&802.6812\\\cdashline{1-3}
Padrón&691.6885&830.0262&Vilariño de Conso&75.6695&90.8034\\\cdashline{4-6}
Rois&6675.1840&8010.2208&Caldas de Reis&2.9690&3.5628\\\cdashline{1-3}
Melide&7789.9880&9347.9856&Cuntis&418.9665&502.7598\\
Santiso&8769.6145&10523.5374&Moraña&20.3405&24.4086\\
Sobrado&5932.7725&7119.3270&Portas&44.9360&53.9232\\
Toques&3208.9905&3850.7886&Valga&170.3230&204.3876\\\cdashline{1-3}\cdashline{4-6}
Vimianzo&11399.9495&13679.9394&Agolada&5022.7920&6027.3504\\
Zas&8707.7510&10449.3012&Dozón&7865.1220&9438.1464\\\cdashline{1-3}
Mazaricos&30523.0320&36627.6384&Lalín&67726.3094&49413.9264\\
Santa Comba&33232.5915&39879.1098&Rodeiro&33535.9704&28899.7230\\\cdashline{1-3}
A Fonsagrada&706.8835&848.2602&Silleda&37476.2754&31263.9060\\
Baleira&5146.0950&6175.3140&Vila de Cruces&7768.5380&9322.2456\\\cdashline{1-3}\cdashline{4-6}
Alfoz&608.9420&730.7304&Meis&14.7590&17.7108\\
Foz&1868.4500&2242.1400&O Grove&13.3965&16.0758\\
Lourenzá&2202.0510&2642.4612&Pontecesures&4.0380&4.8456\\
Mondoñedo&2327.6815&2793.2178&Ribadumia&66.0380&79.2456\\\cdashline{4-6}
O Valadouro&1313.4305&1576.1166&Campo Lameiro&25.3015&30.3618\\\cdashline{1-3}
Ourol&11.3190&13.5828&Cotobade&110.0005&132.0006\\\cdashline{1-3}
A Pontenova&2559.4450&3071.3340&Poio&9.7120&11.6544\\
Barreiros&8867.9365&10641.5238&Pontevedra&153.7085&184.4502\\\cdashline{4-6}
Ribadeo&13780.9415&16537.1298&A Estrada&9735.9840&11683.1808\\
Trabada&8625.3815&10350.4578&Cerdedo&33.4540&40.1448\\\cdashline{1-3}
Antas de Ulla&2370.3995&2844.4794&Forcarei&6233.9230&7480.7076\\
\hline
		\end{tabular}}
		\caption{New systems of milk quotas for $190$ councils in Galicia for the case $\rho=40\%$ with the counties.}\label{rules_190concellos_counties}
	\end{center}
\end{table}

\end{document}